\documentclass[12pt]{article}
\usepackage{amsmath,amsthm,amssymb,epsfig,times}
\usepackage[utf8]{inputenc}
\usepackage[nolists]{endfloat}
\usepackage{subfigure}
\usepackage{natbib}
\usepackage{times}
\usepackage{enumerate}
\usepackage[usenames]{color}
\usepackage{arydshln}

\bibpunct{(}{)}{;}{a}{,}{,}

\newtheorem{theorem}{Theorem}
\newtheorem{definition}{Definition}

\newtheorem{lemma}{Lemma}

\title{\bf Grid-Uniform Copulas and Rectangle Exchanges: Bayesian Model and Inference for a Rich Class of Copula Functions}
\author{{\sc  Nicolás Kuschinski and Alejandro Jara}}
\begin{document}
\date{\today}
\maketitle

\footnotetext[1]{ Nicolás Kuschinski is a Postdoctoral Researcher at the ANID – Millennium Science Initiative Program – Millennium Nucleus Center for the Discovery of Structures in Complex Data, Casilla 306, Correo 22, Santiago, Chile (E-mail: nicolas.kuschinski@mat.uc.cl). Alejandro Jara is 
Associate Professor, Department of Statistics, Pontificia Universidad 
Cat\'olica de Chile and ANID – Millennium Science Initiative Program – Millennium Nucleus Center for the Discovery of Structures in Complex Data Casilla 306, Correo 22, Santiago, Chile (E-mail: 
atjara@uc.cl).  N. Kuschinski's  research is supported by supported by ANID – Millennium Science Initiative Program – NCN17\_059. A. Jara's research is supported by supported by ANID – Millennium Science Initiative Program – NCN17\_059 and Fondecyt 1180640 grant.} 

\begin{abstract}
Copula-based models provide a great deal of flexibility in modelling multivariate distributions, allowing for the specifications of models for the marginal distributions separately from the dependence structure (copula) that links them to form a joint distribution. Choosing a class of copula models is not a trivial task and its misspecification can lead to wrong conclusions. We introduce a novel class of grid-uniform copula functions, which is dense in the space of all continuous copula functions in a Hellinger sense. We propose a Bayesian model based on this class and develop an automatic Markov chain Monte Carlo algorithm for exploring the corresponding posterior distribution. The methodology is illustrated by means of simulated data and compared to the main existing approach.\\
    
\end{abstract}

\section{Introduction}
One of the primary interests of statistical analysis of multivariate data is the study of how random variables relate to each other. Among the various ways to express the relationship between random variables, one of the more flexible ones is the use of a marginals-copula representation, which provides a way to separate multivariate distributions into their single variate marginals and a function which represents their association structure, the copula function \citep{librocop, summarycop, librocop1}.  For any $d$-variate distribution $H:\mathbb{R}^d\rightarrow[0,1]$ with $d>1$ and marginals $F_1,$ $F_2,$ $\ldots,$ $F_d$, a copula is a function $C:\mathbb{R}^d\rightarrow[0,1]$ such that
$$
H(\boldsymbol{x})=C(F_1(x_1),F_2(x_2), \ldots, F_d(x_d)),
$$
where $\boldsymbol{x}=(x_1,\ldots,x_d) \in \mathbb{R}^d$. Sklar's theorem \citep{sklarthm1, sklarthm, librocop} is a classical result which states that this function $C$ exists for any multivariate distribution $H$, and that if $H$ is continuous, $C$ is unique. Copula functions are themselves multivariate probability distributions supported on the unit hyper-cube and are such that the single variate marginals are uniform \citep{librocop}.

Modeling phenomena with copulas involves specifying models for marginals $F_1,$ $F_2,$ $\ldots,$ $F_d$, and for $C(\boldsymbol{x})$ separately. Modeling $F_i(x_i)$ is a matter of modeling single variable probability distributions, which is a  well understood topic and that has been studied with great depth in both frequentist and Bayesian contexts \citep[see, e.g.,][]{bnpbook}. Our primary interest here is the modeling of the copula function $C(\boldsymbol{x})$. There is a large body of literature studying parametric copula models arising from standard multivariate distributions such as the Gaussian and multivariate-$t$  \citep{librocop, bayesgausscop, librocop1, summaryfreqcop, summarycop}. Some popular copula models are members of the Archimedean family, which are single-parameter copulas satisfying
$$
C(u_1,\ldots, u_d\mid \phi, \theta) = \phi^{-1}(\phi(u_1 \mid \theta)+\ldots+\phi(u_d \mid \theta) \mid \theta),
$$
for a function $\phi(\cdot \mid \theta)$ with a single parameter, known as the generator function of the Archimedean copula \cite{librocop}, where $ \phi^{-1}$ is its inverse function. 
Archimedean copulas have been used for both frequentist and Bayesian analyses in the literature \citep{archimcop1, archimcop, bayesarchim}. 

Choosing a class of copula models is not a trivial task and constraining the inference to parametric copula models can lead to wrong conclusions, because it reduces our ability to represent relationships between random variables. Motivated by these facts, different flexible approaches have been discussed in the literature.
Non-parametric approaches for discrete data have been attempted in numerous ways, as described in \cite{countcops} and \cite{discretecops}. In the context of continuous data, classical semiparametric approaches have been discussed by \cite{semiparamfreq}, while classical nonparametric approaches can be traced back to \cite {empcop}, and can be found in \cite{freqnpcop}. The classical methods commonly rely on the use on partial- or pseudo-likelihood and do not allow for a proper quantification of the uncertainties associated to the lack of knowledge of the marginal distributions. Furthermore, these approaches cannot be employed for modelling the association structure of latent variables in the context of hierarchical models. 

Flexible model-based approaches for copula functions can be found 
in \cite{bnpjeffreys}, \cite{gausmixcop}, \cite{skewmixcop}, and \cite{polyacop}. \cite{bnpjeffreys} proposed an interesting semi-parametric Bayesian approach for bivariate copulas based on a finite-dimensional approximation.
This approximation is structured from a partition of the unit interval based of intervals of the same length $\left[(i-1)/m,\,i/m\right]$, where $m$ is the number of intervals and $i\in\{1,\ldots,m\}$. Their proposal is constructed using this partition and indicator functions for the corresponding intervals. The density of the copula is constructed via a mixture of the cross products of the indicator functions, resulting in a locally uniform distribution over the unit square, parameterized by a doubly stochastic matrix. Taking advantage of the properties of doubly stochastic matrices, the authors proceed to develop both a conjugate Jeffreys prior and a Markov chain Monte Carlo (MCMC) algorithm to sample from the posterior distribution. The approach is flexible but is difficult to generalize to not equally-spaced partitions and  higher dimensions because of its reliance on the properties of doubly stochastic matrices. 

\cite{gausmixcop} proposed a model based on mixtures of Gaussian copulas. A Gaussian copula function is given by
$$
C(u_1\ldots u_d)=\Phi_{\boldsymbol{R}}(\Phi^{-1}(u_1),\ldots,\Phi^{-1}(u_d)),
$$
where $\Phi_{\boldsymbol{R}}$ is the CDF of a $d$-variate normal distribution with mean zero, variance one and covariance matrix $\boldsymbol{R}$, arising from the corresponding correlation function, and $\Phi$ is the CDF of a standard normal distribution function. 
The authors claim that mixtures of bivariate Gaussian copulas can approximate any continuous bivariate copula function. Unfortunately, 
the Gaussian copula kernel is not rich enough to form a dense class. For instance, the density function of a bivariate Gaussian copula has the property that $c_{\boldsymbol{R}}(u_1,u_2) = c_{\boldsymbol{R}} (u_2,u_1)$. Therefore, the density of a mixture of arbitrarily many Gaussian copulas also has this feature and, thus, cannot approximate an asymmetrical copula function, such as the  asymmetrical $t$ copula described by
\cite{asymtcop}. \cite{skewmixcop} extended the idea of a mixture of copula functions to the class of skew-normal copulas. However, there is no evidence that this class of copula functions is dense in the space of all copula functions. Finally, \cite{polyacop} employed Dirichlet-based Polya trees models  to propose a fully non-parametric Bayesian approach to modeling copula functions in any number of dimensions and a method for conjugate posterior simulation from the resulting posterior. This attractive result is, however, significantly marred by the flaw that the simulations from the posterior distribution are not themselves copula functions.

In this paper, we introduce a novel class of grid-uniform copula functions, which is dense in the space of all continuous copula functions. We propose a Bayesian model based on this class and develop an automatic MCMC algorithm for exploring the corresponding posterior distribution, allowing for the flexible modelling of continuous joint distributions. The paper is organized as follows. In Section 2 we introduce the class of grid-uniform copula functions and state its main properties, including its ability to approximate any given continuous copula function. In Section 3 we propose a Bayesian model based on the class of grid-uniform copulas and describe the MCMC algorithm.  In Section 4, we illustrate the behavior of the model by means of analyses of simulated data. A final discussion concludes the article.

\section{Grid-Uniform Copulas}

\subsection{Definition}

We begin by defining the class of copulas on which we develop our proposal. Let $\rho$ be an orthogonal grid of $[0,1]^d$. Specifically, let $\rho_i$ be an ordered collection of points in $[0,1]$, and set $\rho = \rho_1 \times \cdots \times \rho_d$, such that $\boldsymbol{1}_d \in \rho$. Let  $\nu^\rho$  be the collection of sets formed by $\rho$, which are indexed by their upper right (or higher dimensional equivalent) corner.  Now, let $F$ be a probability measure defined on an appropriate space and $B$ a measurable set such that $F(B) > 0$. We denote by $F\mid_B$  to the restriction of $F$ to
$B$ defined by $F_{|_B}(A)=F(A\mid B)=F(A \cap B)/F(B)$. A probability distribution $F$ on $[0,1]^d$ is said to be \emph{$\rho$-uniform}, if for each set  $ B \in \nu^\rho$, such that $F(B)>0$, the restriction of $F$ to the set $B$, $F_{|_B}$, is uniform on $B$.  

\begin{definition}{\bf (Grid-uniform copula)}  Let $\rho$ be a grid on $[0,1]^d$. 
    A distribution $C$ on $[0,1]^d$ is a $\rho$-uniform copula if it is $\rho$-uniform and its one-dimensional marginal distributions are uniform.
\end{definition}
A grid-uniform copula can be completely described by specifying the grid $\rho$ and the probabilities for every $B \in \nu^{\rho}$. Hence, for each grid $\rho$, the space of grid-uniform copulas over this grid are a compact finite dimensional domain.  For a grid $\rho$ and a distribution $C$, we will use $C_\rho$ to denote the grid-uniform version of $C$, which assigns to each set $ B \in \nu^\rho$ the probability assigned by $C$ to that set, i.e., $C_\rho(B)=C(B)$ for every $ B \in \nu^\rho$. It is easy to see that if $C$ is a continuous copula, then $C_\rho$ is also a grid-uniform copula and that the CDF of $C_\rho$ and $C$ coincide for every $\boldsymbol{y} \in \rho$.

\subsection{Richness of grid-uniform copulas} \label{hellinger}

We now prove that the class of grid-uniform copulas is sufficiently rich to approximate any arbitrary continuous copula function.

\begin{theorem} Let $C$ be an arbitrary copula which is absolutely continuous with respect to Lebesgue measure. Then for every $\epsilon>0$, there exists a grid-uniform copula $D$ such that the Hellinger distance between $D$ and $C$ is smaller than $\epsilon$, $\mathcal{H}(D,C)<\epsilon$.
\end{theorem}

\noindent {\sc Proof:}  First, we will prove the theorem for the case when $C$ admits a continuous density, denoted by $c$. For each grid $\rho$, let 
$c_\rho$ be the density of the the grid-uniform copula $C_\rho$. Notice now that for every set in $\nu^\rho$, there is a point $\boldsymbol{y}$ such that $c(\boldsymbol{y})=c_\rho(\boldsymbol{y})$. The reason for this is that over the set, $c$ and $c_\rho$ are two continuous functions with the same integral. Notice also that since $c$ is continuous, it is bounded above by some bound $b$. Also, since $c$ is continuous, it is uniformly continuous, i.e., for all points $\boldsymbol{x}$, $\boldsymbol{y}$ and  $\epsilon>0$, there is $\delta >0$ such that if $||\boldsymbol{x}-\boldsymbol{y}||<\delta$, then $|c(\boldsymbol{x})-c(\boldsymbol{y})|\leq\epsilon^2/b$. 

We can now pick a grid $\rho^\star$ such that for all points $\boldsymbol{x}$, $\boldsymbol{y}$ within the same set of $\rho\star$,  $||\boldsymbol{x}-\boldsymbol{y}||<\delta$. By the above observation, it follows that $|c_{\rho^\star}(\boldsymbol{x})-c(\boldsymbol{x})|=|c_{\rho^\star}(\boldsymbol{y})-c(\boldsymbol{x})|=|c(\boldsymbol{y})-c(\boldsymbol{x})|<\epsilon^2/b$. It follows that  the squared Hellinger distance between $C$ and $C_{\rho^\star}$, is given by
\begin{eqnarray}\nonumber
    \mathcal{H}^2(C, C_{\rho^\star}) &= &1-\int_{[0,1]^d} \sqrt{c(\boldsymbol{x}) c_{\rho^\star}(\boldsymbol{x})}d\boldsymbol{x},\\ \nonumber
            &= &1-\int_{[0,1]^d}\sqrt{c(\boldsymbol{x})(c_{\rho\star}(\boldsymbol{x})-c(\boldsymbol{x}))+c(\boldsymbol{x}))}d\boldsymbol{x}, \\\nonumber
            &\leq & 1-\int_{[0,1]^d} c(\boldsymbol{x}) d\boldsymbol{x} + \int_{[0,1]^d}\sqrt{c(\boldsymbol{x})|c_{\rho^\star}(\boldsymbol{x})-c(\boldsymbol{x})|}d\boldsymbol{x}, \\ \nonumber
            &\leq & \sqrt{b\epsilon^2/b}=\epsilon. \\\nonumber
\end{eqnarray}
For the case when $C$ does not admit a continuous density, we proceed with a somewhat similar idea. Let $c$ be a density of $C$.
Let  $E$ be the set of discontinuities of $c$. For every grid $\rho$, let $\nu^\rho_1$ be the collection of sets which have a discontinuity and $\nu^\rho_2$ be the collection of sets which do not have a discontinuity. Now, there is a grid $\rho_1$ such that the measure that $C$ assigns to $\nu^{\rho_1}_1$ is less than $\epsilon/3$, for any $\epsilon>0$. Furthermore, for every set in $\nu^{\rho_1}_2$, $C$ is uniformly continuous and has an upper bound $b$, so by the same argument used in the case when $c$ was continuous, we can find a new grid $\rho_2$ such that $|c_{\rho_2}(\boldsymbol{x})-c(\boldsymbol{x})|<\epsilon^2/9b$ for all $\boldsymbol{x}$ contained in one of the sets of $\nu^\rho_{2}$.

Now let $\rho^\star$ be any grid which is a refinement of $\rho_1$ and $\rho_2$.  It follows that  the squared Hellinger distance between $C$ and $C_{\rho^\star}$, is given by 
\begin{eqnarray}\nonumber
    \mathcal{H}^2(C,C_{\rho^\star})&=&\frac{1}{2}\int_{[0,1]^d}\left(\sqrt{c(\boldsymbol{x})}-\sqrt{c_{\rho^\star}(\boldsymbol{x})}\right)^2 d\boldsymbol{x}, \\\nonumber
            &=&\frac{1}{2}\left(\sum_{k_i \in \nu^{\rho_1}_1}\int_{k_i}\left(\sqrt{c(\boldsymbol{x})}-\sqrt{c_{\rho^\star(\boldsymbol{x})}}\right)^2 d\boldsymbol{x} +\sum_{k_i \in \nu^{\rho_1}_2}\int_{k_i}\left(\sqrt{c(\boldsymbol{x})}-\sqrt{c_{\rho^\star}(\boldsymbol{x})}\right)^2 d\boldsymbol{x}\right), \\\nonumber
            &\leq&\epsilon/3 + \frac{1}{2} \sum_{k_i \in \nu^{\rho_1}_2} \int_{k_i} \left(\sqrt{c(\boldsymbol{x})} - \sqrt{c_{\rho^\star} (\boldsymbol{x})} \right)^2 d\boldsymbol{x}, \\\nonumber
            &\leq&\epsilon/3 + 1-\left(\sum_{k_i \in \nu^{\rho_1}_2}\int_{k_i}\sqrt{c(\boldsymbol{x}) c_{\rho^\star}(\boldsymbol{x})} d\boldsymbol{x}\right),\\\nonumber
            &=&\epsilon/3 + 1-\left(\sum_{k_i \in \nu^{\rho_1}_2} \int_{k_i} \sqrt{c(\boldsymbol{x}) (c_{\rho\star}(\boldsymbol{x})-c(\boldsymbol{x})) + c(\boldsymbol{x}))} d\boldsymbol{x}\right), \\\nonumber
            &=&\epsilon/3 + 1-\sum_{k_i \in \nu^{\rho_1}_2} \left(\int_{k_i} c(\boldsymbol{x}) d\boldsymbol{x} + \int_{k_i}\sqrt{c(\boldsymbol{x}) |c_{\rho^\star}(\boldsymbol{x})-c(\boldsymbol{x})|} d\boldsymbol{x} \right), \\\nonumber
            &\leq&\epsilon/3 + \epsilon/3 +\sqrt{b\epsilon^2/9b}=\epsilon,\\\nonumber
\end{eqnarray}
which completes the proof of the theorem.\hfill$\square$ 

We may note that none of the steps in the proof actually make use of the fact that $C$ has uniform marginals, and it can easily be extended to prove that grid-uniform distributions can approximate any continuous distribution over a rectangular support.

\subsection{Measures of association}
In this section we provide the expression for two important measures of dependence for grid-uniform copula models. Specifically, we provide the expression for Kendall’s tau and Spearman’s rho, which are considered the best alternatives to the linear correlation coefficient as a measure of dependence for non-elliptical distributions, for which the linear correlation coefficient is inappropriate and often misleading. 

Let $C$ be a $\rho$-uniform copula function. 
Let $C^{(i,j)}$ be the bivariate marginal copula of $C$  for the variables in the coordinates $i$ and $j$. 
Let $a_{(i,k)}$, $k=0,\ldots, m_i$, be the $k$th element in $\rho_i$ and $b_{(j,l)}$, $l=0,\ldots, m_j$, be the $l$th element in $\rho_j$. 
It is straightforward to show that
$C^{(i,j)}$ is a $\rho^{(i,j)}$-uniform copula function, where  $\rho^{(i,j)}=\left(\rho_{i},\rho_{j} \right)$ and
$$
C^{\left(i,j\right)}\left ( \left( a_{(i,k-1)},a_{(i,k)} \right]\times \left(b_{(j,l-1)},b_{(j,l)}\right] \right)  =   \sum_{B\in S_{i,j}^{\left(a_{(i,k)},b_{(j,l)}\right)}}C \left( B \right),
$$
where $S_{i,j}^{\left(a_{(i,k)}, b_{(j,l)}\right)}$ is the collection of sets in $\nu^\rho$ such that $i$th coordinate of the index of the set is $a_{(i,k)}$ and the $j$th coordinate of the index of the set is $b_{(j,l)}$.  Spearman’s rho, $\beta$, and  Kendall’s tau, $\tau$, for  the variables in the coordinates $i$ and $j$ is given by
$$
\beta=3\sum_{k=1}^{m_i}\sum_{l=1}^{m_j} \left(a_{(i,k)}^2-a_{(i,k-1)}^2 \right) \left(b_{(j,l)}^2-b_{(j,l-1)}^2 \right) c_{k,l} - 3,
$$
and
$$
\tau=4\sum_{k=1}^{m_i}\sum_{l=1}^{m_j} \left(a_{{i,k}}^2-a_{(i,k-1)}^2\right) \left(b_{(j,l)}^2-b_{(j,l-1)}^2\right) c_{k,l}^2 - 1,
$$
respectively, where 
$$
c_{k,l}= \frac{ C^{\left(i,j\right)}\left ( \left( a_{i,k-1},a_{i,k} \right]\times \left(b_{i,l-1},b_{(i,l)}\right] \right) } {\left(a_{(i,k)}-a_{(i,k-1)} \right) \left( b_{(j,l)}-b_{(j,l-1)} \right)}.
$$

\subsection{Rectangle exchanges}

We now introduce a class of transformations on grid-uniform copulas referred to as rectangle exchanges, which have the following important properties: (i) a rectangle exchange on a $\rho$-uniform copula produces another $\rho$-uniform copula, and (ii) given a grid $\rho$, and two $\rho$-uniform copulas $C$ and $D$, there is a finite sequence of rectangle exchanges which can transform $C$ into $D$.

\begin{definition} Let $\rho$ be a grid on $[0,1]^d$ and  $C$ be a grid $\rho$-uniform copula function. The function $C^*$ is the result of a rectangle exchange of  $C$,
if  $C^*$ is constructed using the following steps:
\begin{itemize}
    \item[(1)] Set $C^* = C$, and pick $i$ and $j$ in the set $\{1,\ldots, d\}$, such that $i <j$ and the cardinality of $\rho_i$ and $\rho_j$ is greater than or equal to 2. Also, for all $k \in \{1,\ldots,d\} \setminus \{i,j\}$ pick point $x_k \in \rho_k$.
    
    \item[(2)] Pick $a_1,a_2 \in \rho_i$ and $b_1,b_2 \in \rho_j$. 

    \item[(3)] Set 
    $$
    \boldsymbol{p}_{(a_l,b_m)} = (x_1,\ldots,x_{i-1},a_l,,x_{i+1},\ldots,x_{j-1},b_m, x_{i+1}, \ldots,x_d), 
    $$
    where $l,m \in \{1,2\}$.  
        
    \item[(4)] Pick some $\epsilon$ in the interval 
        $$\left[\max\left\{-C\left(\nu^{\rho}_{\boldsymbol{p}_{(a_1,b_2)}} \right) ,-C\left(\nu^{\rho}_{\boldsymbol{p}_{(a_2,b_1)}} \right) \right\}, 
                \min\left\{ C\left(\nu^{\rho}_{\boldsymbol{p}_{(a_1,b_1)}} \right),C\left(\nu^{\rho}_{\boldsymbol{p}_{(a_2,b_2)}} \right) \right\}
        \right].$$
        
            \item[(5)] Set
        \begin{itemize}
            \item[(a)] $C^*\left(\nu^{\rho}_{\boldsymbol{p}_{(a_1,b_1)}} \right) = C\left(\nu^{\rho}_{\boldsymbol{p}_{(a_1,b_1)}} \right)  - \epsilon,$
            \item[(b)] $C^*\left(\nu^{\rho}_{\boldsymbol{p}_{(a_1,b_2)}} \right) = C\left(\nu^{\rho}_{\boldsymbol{p}_{(a_1,b_2)}} \right) + \epsilon,$
            \item[(c)] $C^*\left(\nu^{\rho}_{\boldsymbol{p}_{(a_2,b_1)}} \right) = C\left(\nu^{\rho}_{\boldsymbol{p}_{(a_2,b_1)}} \right) +\epsilon,$
            \item[(d)] $C^*\left(\nu^{\rho}_{\boldsymbol{p}_{(a_2,b_2)}}\right) =  C\left(\nu^{\rho}_{\boldsymbol{p}_{(a_2,b_2)}}\right)  - \epsilon.$
        \end{itemize}
\end{itemize}
\end{definition}

\begin{figure}
\centering
\subfigure[Step (1)] 
{
    \includegraphics[width=7.5cm]{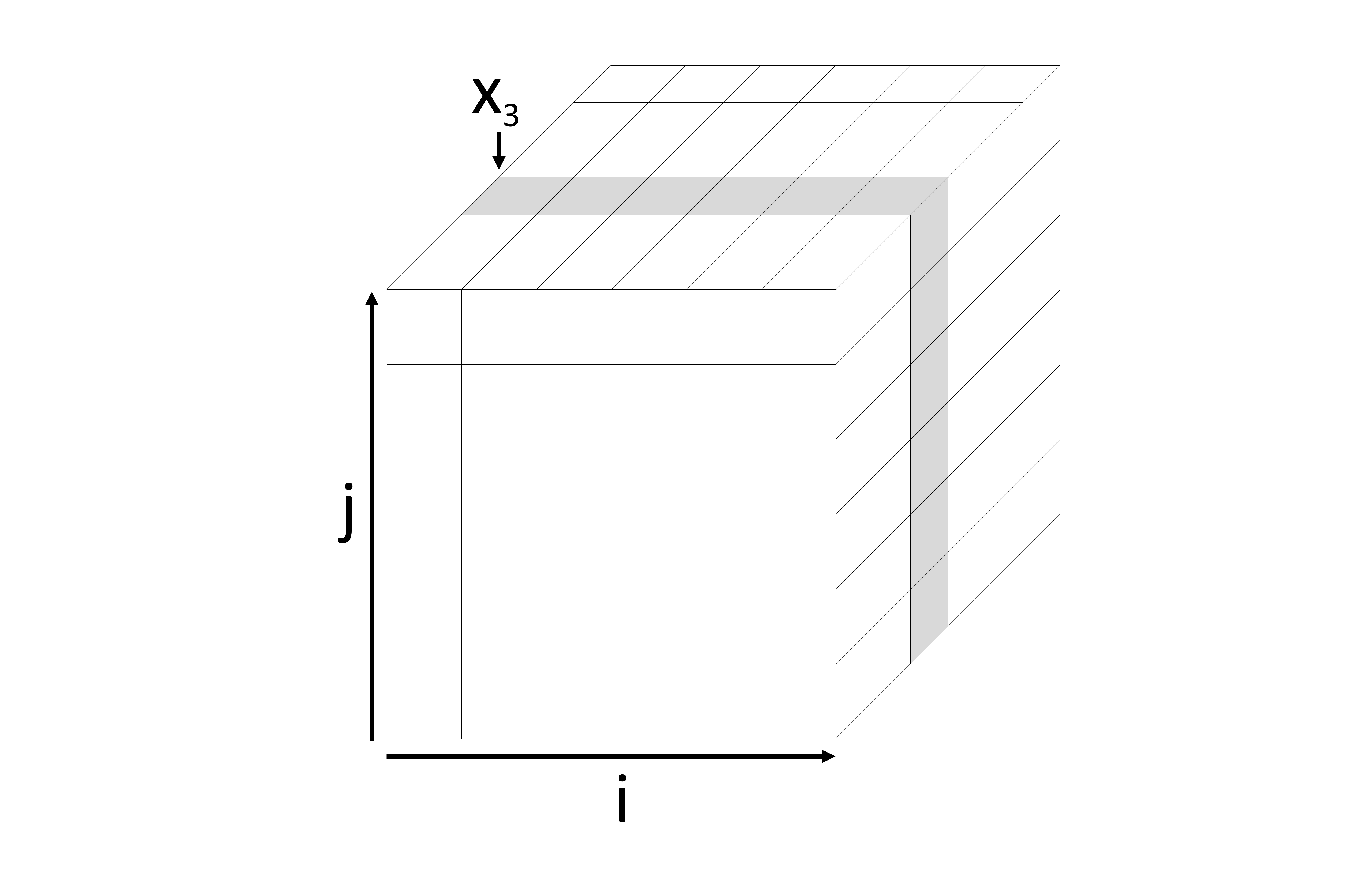}
}
\subfigure[Step (2)] 
{
    \includegraphics[width=7.5cm]{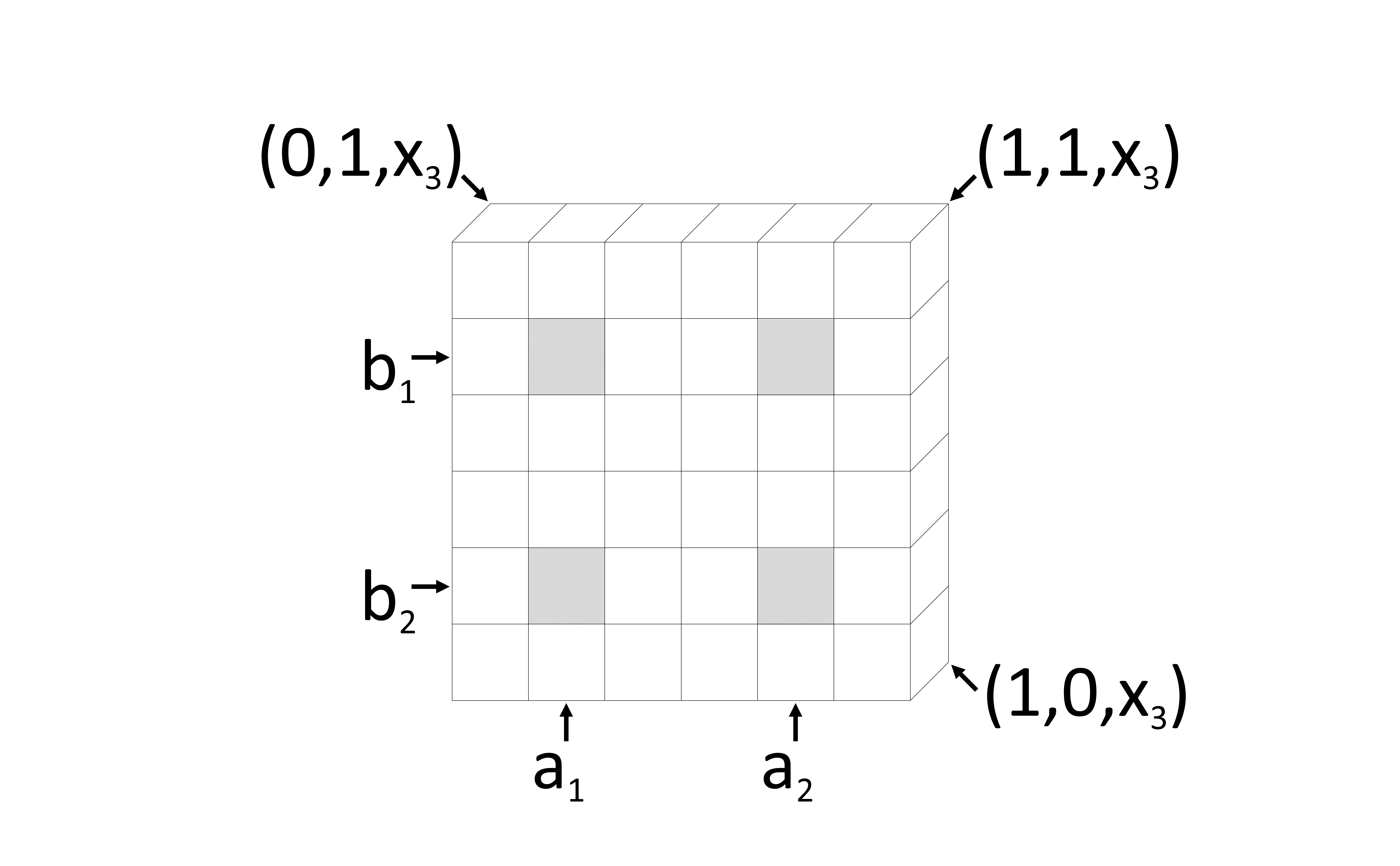}
}\\
\subfigure[Step (3)] 
{
    \includegraphics[width=7.5cm]{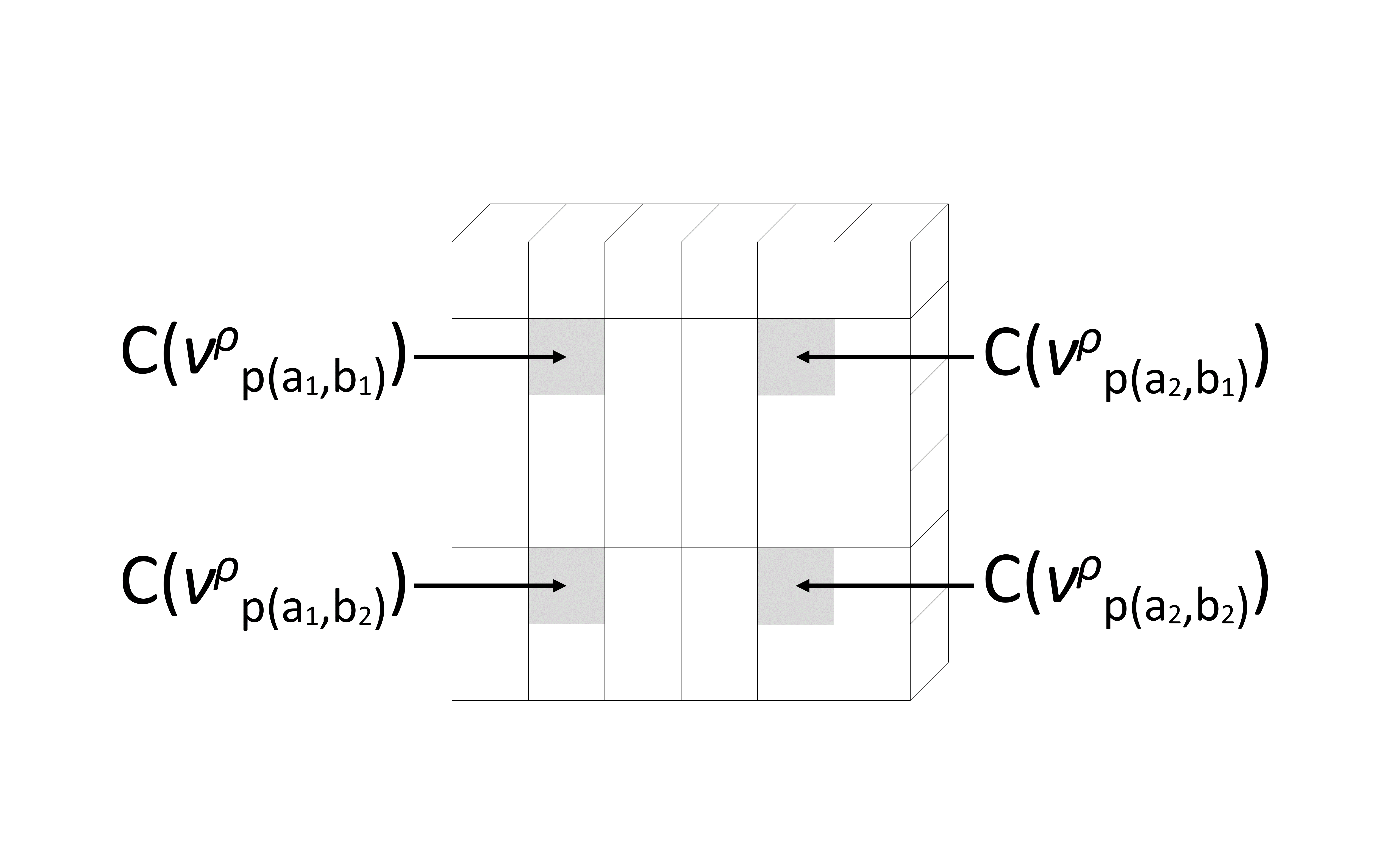}
}
\subfigure[Step (5)] 
{
    \includegraphics[width=7.5cm]{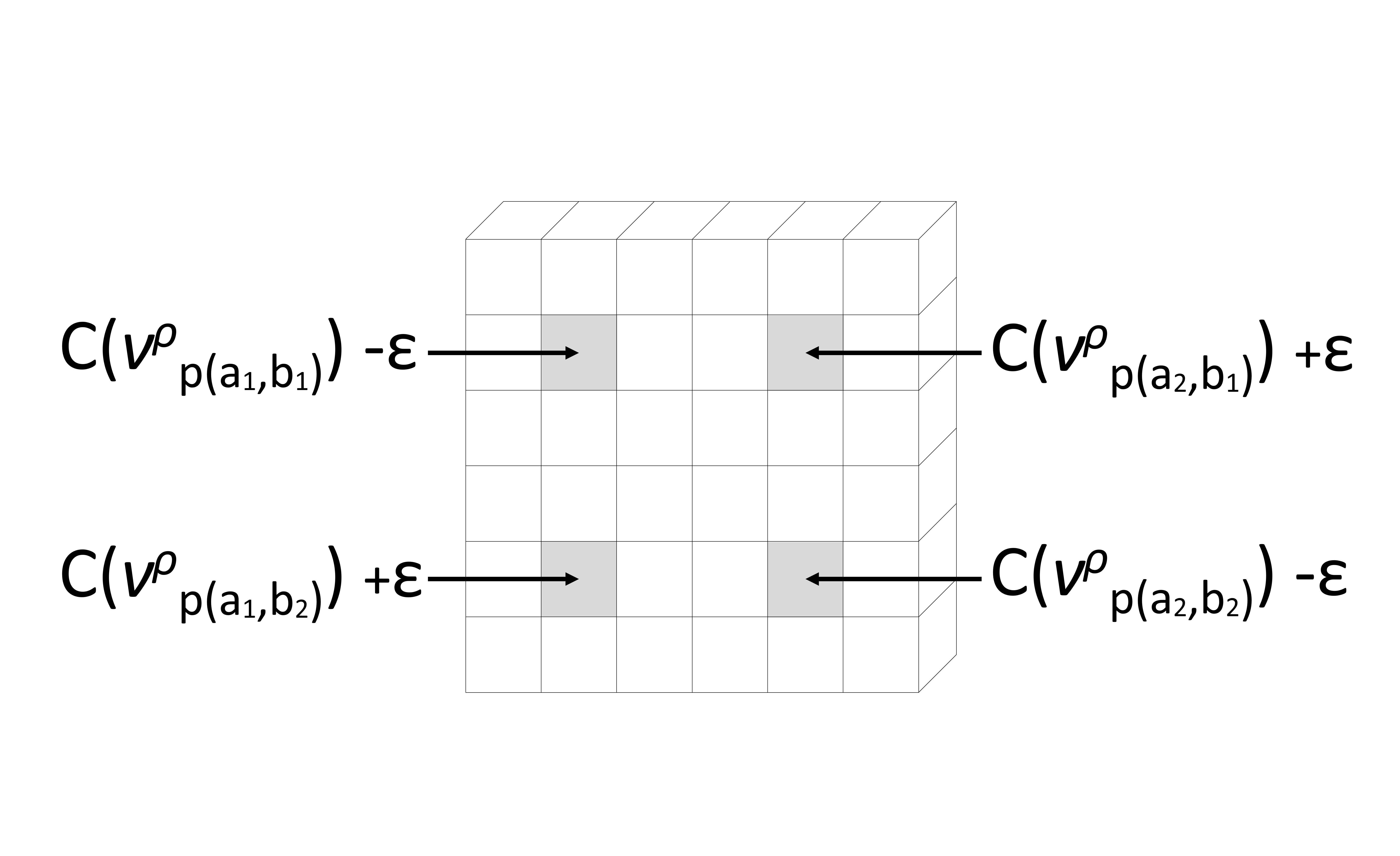}
}
\caption{\label{figure:rectangleswitch} Rectangle exchange - Illustration of a rectangle exchange for a $\rho$-uniform bivariate copula function $C$, where
    $\rho_1$ and $\rho_2$ have 7 equally-spaced points. Panel (a) illustrates step (1), where $i$ and $j$ are picked from $\{1,\ldots,d\}$, and $x_3$ is picked from $\rho_3$. Panel (b) illustrates step (2), where $i,j \in \{1,\ldots, d\}$ and $b_1,b_2 \in \rho_j$ are picked and where the rectangles to be exchanged are selected. Panel (c) illustrates step (3), showing the assigned sets and the mass assigned to them by $C$. Panel (d) illustrates step (5), where  the mass of the new copula function $C^*$ is computed.}
\end{figure}

The rectangle exchange operation is illustrated in two-dimensions in Figure~\ref{figure:rectangleswitch}.

Rectangle exchanges are a closed operation on grid-uniform copulas since they conserve the uniformity of all marginals.
\begin{lemma} \label{cerradura} Let $G$ be $\rho$-uniform copula and $\mathcal{G}$ the resulting rectangle exchange of $G$, Then $\mathcal{G}$ is also a $\rho$-uniform copula.
\end{lemma}
\noindent {\sc Proof:}  We will call the two coordinates of the two dimensional plane on which the exchange was performed, coordinate $i$ and coordinate $j$, respectively. For considering marginals of coordinates other than $i$ and $j$, we observe that all slices contain either no changed set, or all of the changed sets. For slices without any changed set, we note that the probability has not changed, and for slices with all of the changed sets, the magnitude of the change is $\epsilon - \epsilon + \epsilon - \epsilon=0$, so the marginals along these coordinates remain uniform.

For coordinate $i$ and $j$, we can assume, without loss of generality, that we are only worried about coordinate $i$, and that $i$ is the first coordinate in the two dimensional slice on which the rectangle exchange was performed. Then slices in direction $i$ come in two varieties: slices other than the ones along $a_1$ and $a_2$ contain no changed sets, so the probability of the slice remains unchanged. The slice along $a_1$ contains the sets $\nu^\rho_{\boldsymbol{p}_(a_1,b_1)}$ and $\nu^\rho_{\boldsymbol{p}_(a_1,b_2)}$ so the probability of the slice changes by $-\epsilon+\epsilon=0$, and slices along $a_2$ are similar. Hence, rectangle exchanges conserve uniformity for all marginals.
\hfill$\square$ 

Another interesting property of grid-uniform copula functions is that starting from an uniform distribution on $[0,1]^d$, it is possible to reach any given grid-uniform copula, say $C_0$, by doing certain type of operations. Specifically for a given grid-uniform copula $C$, we will refer as a \textit{grid division} on $C$, to the addition of a division along any of the coordinates, such that the sets that are not divided retain their probabilities, and those that are divided distribute their probability in proportion to their volume. In other words, the resulting copula function arising from a grid division of $C$ is identical to $C$, but is mapped onto a more refined grid.

\begin{lemma} \label{completez} Let $U$ be a uniform distribution in $[0,1]^d$, and let $C$ be an arbitrary grid-uniform copula function. Then, there is a finite sequence of grid divisions and rectangle exchanges which will transform $U$ into $C$.
\end{lemma}
\noindent {\sc Proof:}  We will prove this for the two-dimensional case. For higher dimensions, the proof is identical, but the notation is more complex. We proceed by induction on the size of the grid. For $C$ on grids up to $2\times 2$ the result is obvious, in particular, on any grid of size $1\times d$, the only grid-uniform copula is a completely uniform distribution. Now we assume that it is possible to transform $U$ into $Q$ for all grid-uniform copulas $Q$ on grids of size $m\times k$.  Assume that $C$ is a grid-uniform copula on a grid of size $m\times (k+1)$ and we will prove that it is possible to transform $Q$ into $C$. It is also necessary to consider the case of splitting the other coordinate, that is, the case where $C$ is $(m+1)\times k$. However,  the proof is identical.

Consider the distribution $C'$ such that $C'(A)=C(A)$ for $A\subseteq\bigcup_{i}\bigcup_{j<k} \nu_{ij}$ and where the last two sets of each row have had the probability distributed uniformly between them. We note that $C'$ is a grid-uniform copula on an $m\times k$ grid. We will now show a sequence of steps, starting at $C'$ which will lead to $C$. Begin by splitting the last column along the last division in the grid of $C$, thus making the grids equal. Now we have a distribution which has the same grid as $C$ and which is equal except for $2m$ sets. We will update $C'$ to be this new distribution.  We note that for all $i$, $C'(\nu^\rho_{i,k})+C'(\nu^\rho_{i,k+1})=C(\nu^\rho_{i,k})+C(\nu^\rho_{i,k+1})$. Hence, if the only transformations performed are rectangle exchanges, and if all of those that include $\nu^\rho_{i,k}$ also include $\nu^\rho_{i,k+1}$ then this sum will remain constant.  It can, therefore, be concluded that if only rectangle exchanges of this kind are used, it is sufficient to adjust the probability of the $k$th column.

Now we will label the number of sets in the $k$th that are different in $C'$ and $C$ as $q$. Observe that the sum of the column is the same in both $C'$ and $C$ because they are copulas. Therefore, if $C'\neq C$ at least one set must have a probability which is higher than the probability of $C$ and one set must have a lower probability.
Pick the index of any of the sets with higher probability and call it $\alpha$, and any of the lower sets and call it $\beta$. Now perform a rectangle exchange with $a_1=\alpha$, $a_2=\beta$, $b_1=k$, $b_2=k+1$ and $\epsilon=\min\{C'(\nu^\rho_{\alpha,k})-C(\nu^\rho_{\alpha,k}), C(\nu^\rho_{\beta,k})-C'(\nu^\rho_{\beta,k})\}$. This will yield a new grid-uniform copula with at most $q-1$ probabilities in the $k$th column different from those in $C$. The proof is completed by iterating this procedure as necessary.
\hfill$\square$ 

These results allow us to prove that via rectangle exchanges is possible to generate the full space of grid-uniform copula functions.

\begin{theorem} The $C_1$ and $C_2$ be two $\rho$-uniform copulas. There is a finite sequence of rectangle exchanges to transform $C_1$ into $C_2$.
\end{theorem}

\noindent {\sc Proof:}  Lemma \ref{cerradura} proves that if a distribution can be created by performing a rectangle exchange then it is a $\rho$-uniform copula.
Lemma \ref{completez} shows us that it is possible to reach an arbitrary $\rho$-uniform copula, $C_0$, from a Uniform distribution by means of rectangle exchanges and grid-divisions; however the grid divisions are not actually necessary. To see this, note that the uniform distribution $U$ is already a $\rho$-uniform copula.
We can follow the procedure from the previous lemma if we consider $U$ in it's $\rho$-uniform representation.
The difference is that some rectangle exchanges in the procedure described previously were performed between sets which are the union of several elements of $\nu^\rho$.
At the time of the exchange, these unions of sets can be seen as a single superset with probability distributed uniformly.
We can achieve the result of the larger exchange by performing exchanges with the component sets, adjusting the exchanged probability ($\epsilon$) in proportion to the volume of the set, the details of exactly how to do this are explained next.

Let $A_l$, $A_k$, $A_m$ and $A_n$ be the sets involved in a larger rectangle exchange and let $A_{i}=\bigcup A_{i,j}$ sets which result from adding grid divisions.
For any set $A_i$, we will refer to its probability before the exchange as $P_i$ and its probability after the exchange as $Q_i$.
We proceed to describe how this same probability distribution of the rectangle exchange on $A_i$ can be achieved by performing smaller rectangle exchanges on the sets in $A_{i,j}$.

We will consider the case where a single additional grid division is performed. In this case exactly two sets, $A_i$ are split.
With no loss of generality, we can assume that $A_l$ and $A_k$ were split. 
Hence $A_l=A_{l,1}\cup A_{l,2}$ and $A_k=A_{k,1}\cup A_{k,2}$.
Note that $Q_l=P_l+\zeta$ and $Q_k=P_k-\zeta$ for $\zeta=\pm\epsilon$.
Note also that $P_{l,1}=\xi P_{l}$ and $P_{k,1}=\xi P_{k}$ where $\xi$ is the volume of $A_{l,1}$ divided by the volume of $A_l$ (and matches with $A_k$ and $A_{k,1}$).
Hence $Q_{l,1}=P_{l,1}+\xi\zeta$ and $Q_{k,1}=P{k,1}-\xi\zeta$.
We therefore perform a rectangle exchange with $A_m, A_n, A_{l,1},$ and $A_{k,1}$ exchanging the probability $\xi\zeta$ and then another with $A_m, A_m, A_{l,2},$ and $A_{k,2}$ exchanging the probability $(1-\xi)\zeta$.
Thus, we have shown how a single rectangle exchange can be emulated by performing a rectangle exchange over a grid with an additional split.
For refinements of the grid which add more than one additional split, add the splits one at a time.

The remaining issue is to prove that if $C_0$ and $C_1$ are both $\rho$-uniform copulas, then it is possible to perform rectangle exchanges to get from $C_0$ to $C_1$.
One way to do this is to reverse the steps to arrive at $C_0$ from a uniform copula, and then go from the uniform copula to $C_1$, which completes the proof of the theorem.
\hfill$\square$ 

There is something surprising going on here. Intuition would lead us to believe that rectangle exchanges would work for 2 dimensions, whereas higher dimensions would require parallelepiped exchanges. However, the surprising fact is that the above theorem is true regardless of the number of dimensions. In essence, the apparently very complex problem of exploring the space of grid-uniform copulas is solved in any number of dimensions by repeated transformations of the sort illustrated in Figure~\ref{figure:rectangleswitch}.

\section{Bayesian modeling and inference using grid-uniform copulas}

Our ultimate objective is to use grid-uniform copulas to perform Bayesian statistical inference. We state the components of the Bayesian model in this section.
Assume that we observe an independent and identically distributed (i.i.d.) sample of size $n$ from a $d$-variate continuous distribution $H$, $\boldsymbol{y}_1, \ldots, \boldsymbol{y}_n \mid H \overset{i.i.d.}{\sim} H$, 
where  $H(\boldsymbol{y})=C(F_1(y_1),F_2(y_2), \ldots, F_d(y_d))$, with $F_1,\ldots, F_n$ being the marginal distributions of $H$, and $C$ is the corresponding copula function. We model $C$ as a grid-uniform copula function. Under the grid-uniform copula model, the log-likelihood function is given by:
\begin{eqnarray}\label{like}\nonumber
\ell(C, F_1,\ldots,F_d \mid \boldsymbol{y}_1,\ldots,  \boldsymbol{y}_n) &=& \sum_{i=1}^n \sum_{j=1}^d  \log \left(f_j\left(y_{ij}\right) \right) +  \\
&& \sum_{i=1}^n  \sum_{j=1}^{\left|\nu^{\rho} \right|} \log\left( \frac{C_{\rho} (B_j)}{\lambda\left(B_j \right)}\right) \times I_{\left \{ \left(F_1(y_{i1}), \ldots, F_d(y_{id}) \right) \in B_j  \right \}}\left(\boldsymbol{y}_i\right),
\end{eqnarray}
where $B_j \in \nu^{\rho}$, $\left|\nu^{\rho} \right|$ is the cardinality of $\nu^{\rho}$, $\lambda\left(B \right)$ is the Lebesgue measure of the set $B$, $I_A(B)$ is the indicator function that takes the value $1$ if $B \in A$, and $0$ otherwise.

\subsection{Grid-uniform prior models} \label{priors}

Let $\rho$ be a grid on $[0,1]^d$.  Let $C_0$ be an arbitrary reference copula function and $\alpha > 0$. We propose prior models for grid-uniform copula functions of the form
$$
\pi(C \mid \rho, \alpha, C_0 )\propto \exp\left\{ -\frac{1}{2} \alpha \times \mathcal{D}\left(C,C_0 \right)\right\} \times I_{\mathcal{C}_{\rho}}(C),
$$
where $\mathcal{D}$ be a suitable distance for probability distributions, and $\mathcal{C}_{\rho}$  is the space of $\rho$-uniform copulas. Many choices for $\mathcal{D}$ could be considered. One choice that provides a simple interpretation of the hyper-parameters is the squared-$L_2$ distance. Let $c_0$ and $c$ be densities for $C_0$ and $C$, respectively. Let $B_1, \ldots, B_p$ be the sets included in $\nu^\rho$.  Under the squared-$L_2$ distance the grid-uniform prior model is given by 
\begin{eqnarray}\nonumber  \label{prior1}
 \pi(C \mid \rho, \alpha, C_0 ) &\propto& \exp \left\{  -\frac{\alpha}{2}  \times \int_{[0,1]^d} \left(c(\boldsymbol{x}) - c_0(\boldsymbol{x}) \right)^2 d \boldsymbol{x} \right\}  \times I_{\mathcal{C}_{\rho}}(C),\\\nonumber
     &=& \exp \left\{ -\frac{\alpha}{2}  \times 
    \sum_{l=1}^{\left|\nu^{\rho} \right|} \left[  \int_{B_l} (c(\boldsymbol{x})-c_0(\boldsymbol{x}))^2 d \boldsymbol{x}  \right] \right\} \times I_{\mathcal{C}_{\rho}}(C),\\
        &\propto& \exp \left \{ -\frac{\alpha}{2}  \times \sum_{l=1}^{\left|\nu^{\rho} \right|} \left[-2c_l \int_{B_l} c_0(\boldsymbol{x}) d \boldsymbol{x}  + c_l^2 \right] \right\} \times I_{\mathcal{C}_{\rho}}(C),
\end{eqnarray}
where $c_l= \frac{\int_{B_l} c(\boldsymbol{x}) d \boldsymbol{x} }{\lambda (B_l)}$,  with $\lambda(A)$ being the  Lebesgue measure of the set $A$. The prior takes the form of a truncated $\left|\nu^{\rho} \right|$-variate normal random distribution, centered at the $\rho$-uniform version of $C_0$, $C_{0,\rho}$, and precision matrix given by  $\alpha \times \mathbf{I}_{\left|\nu^{\rho} \right|}$. 

\begin{figure}
\centering
\subfigure[$\alpha^\star=2$.] 
{
    \includegraphics[width=7cm]{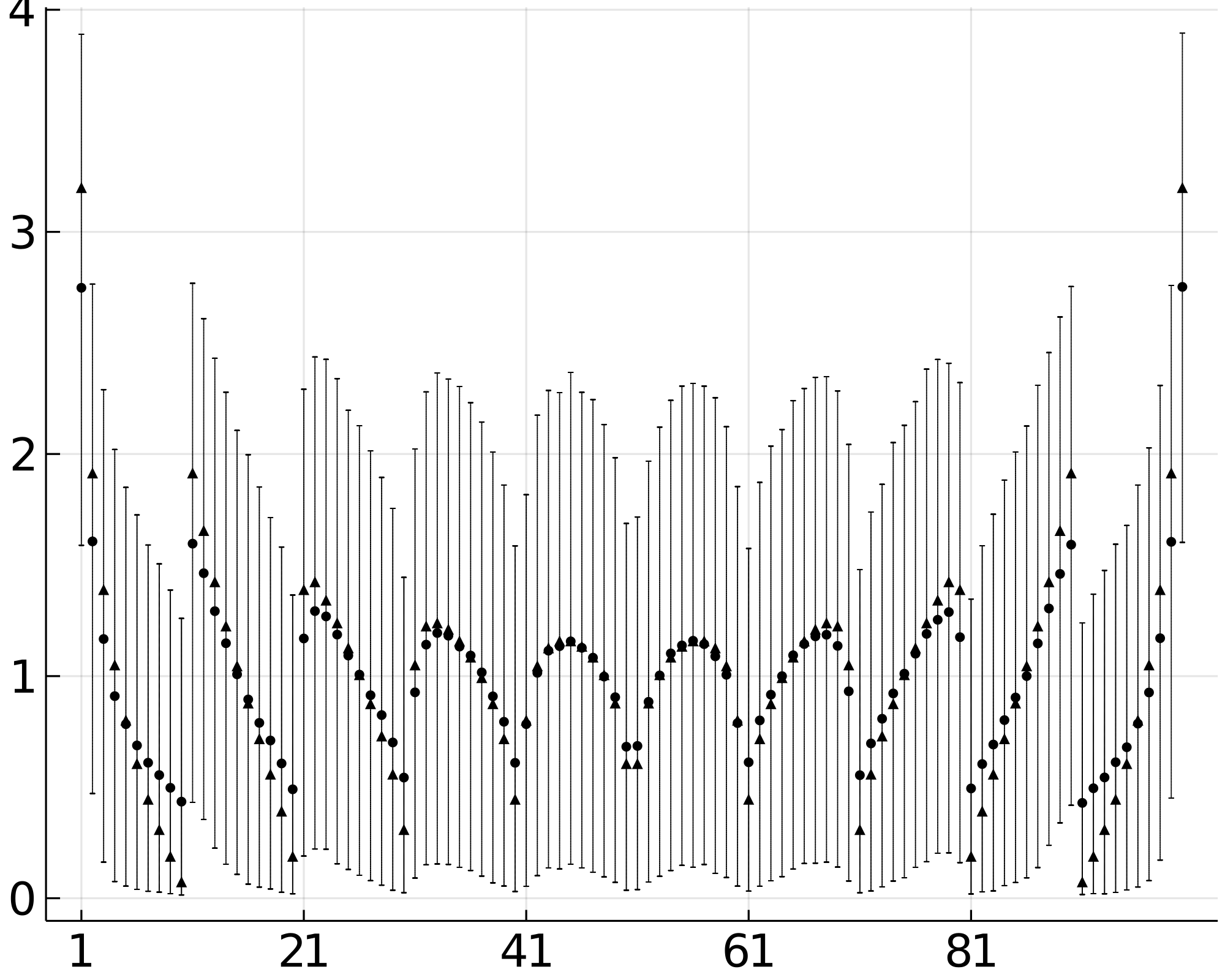}
}
\subfigure[$\alpha^\star=20$.] 
{
    \includegraphics[width=7cm]{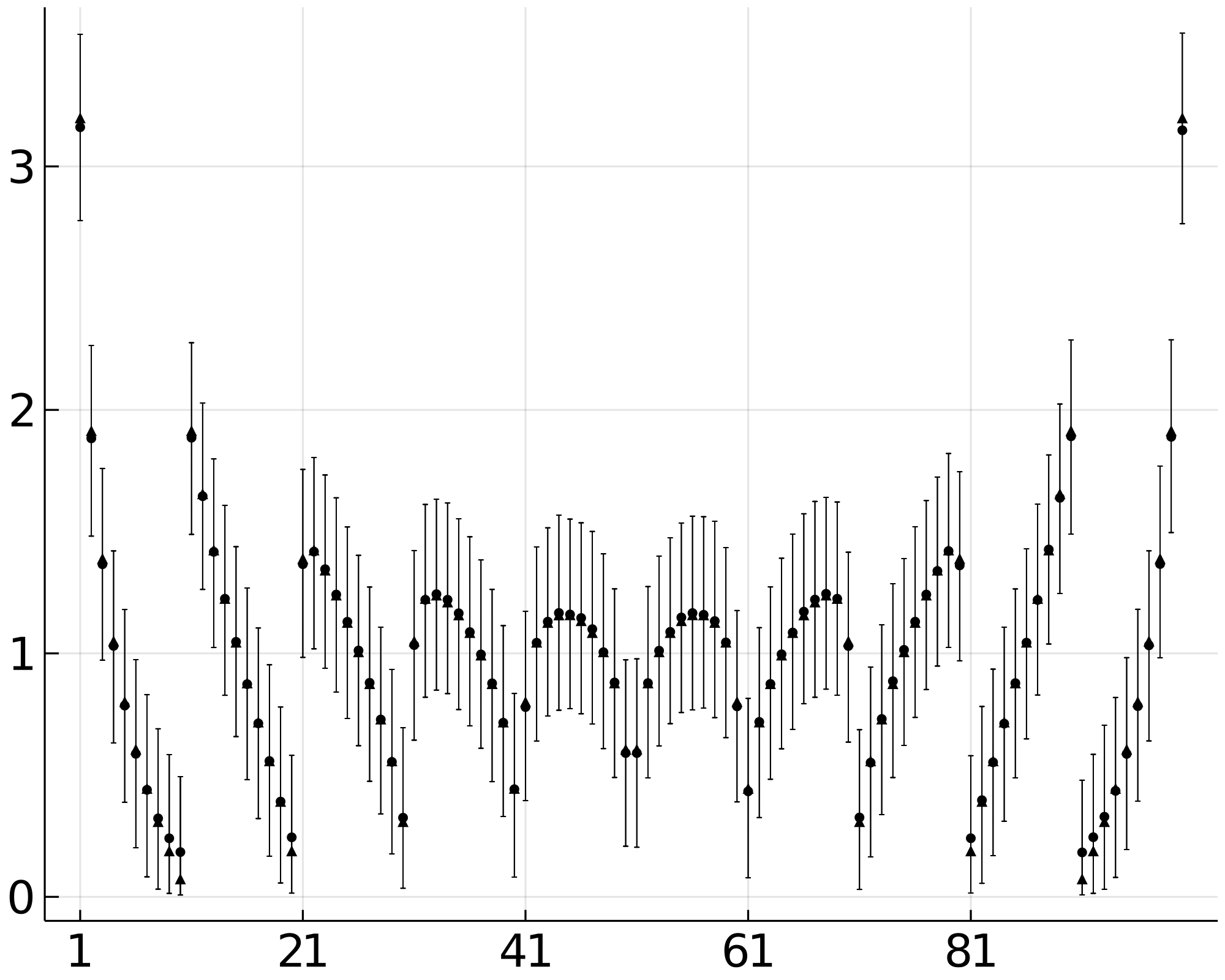}
}\\
\subfigure[$\alpha^\star=200$.] 
{
    \includegraphics[width=7cm]{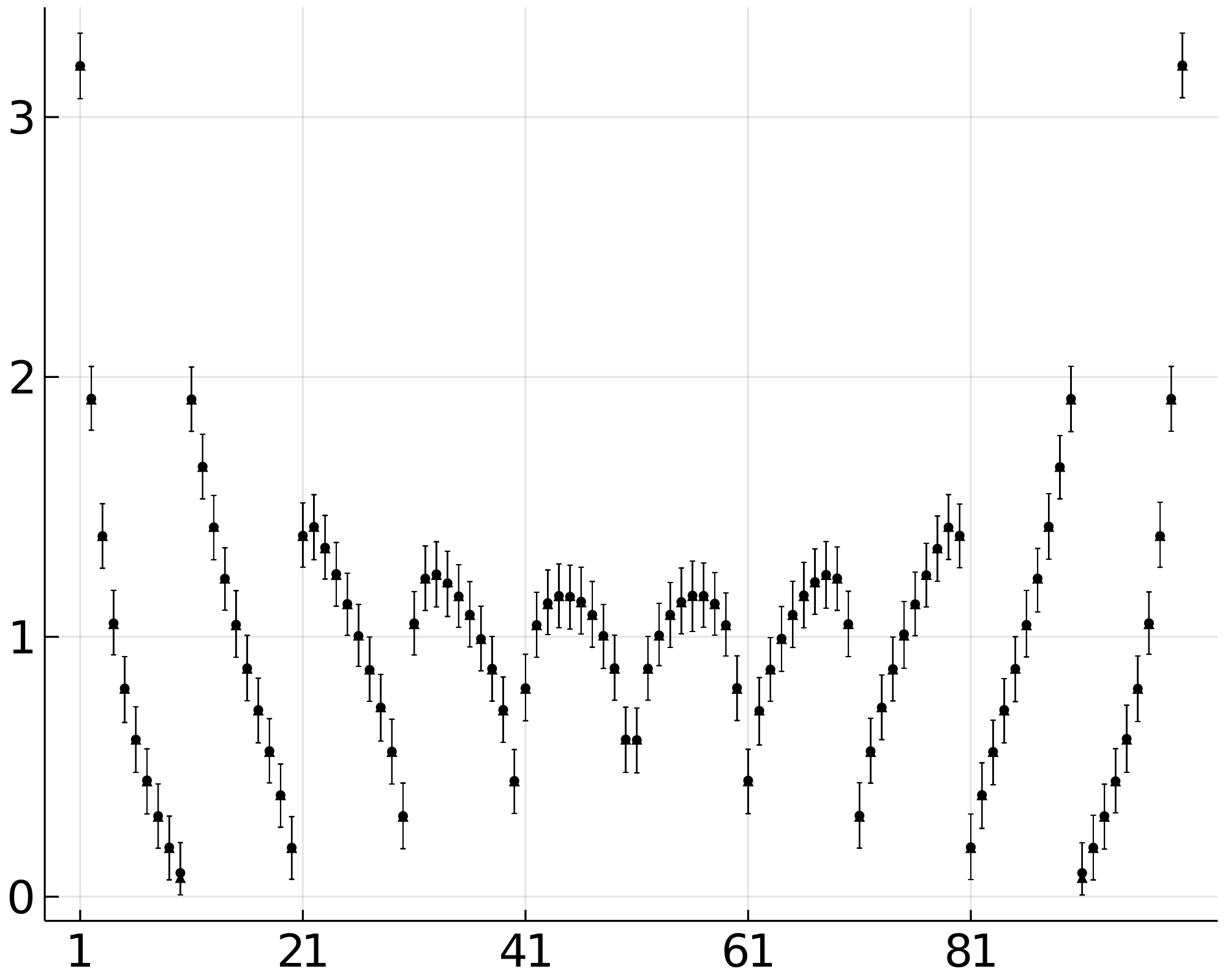}
}
\subfigure[$\alpha^\star=2000$.] 
{
    \includegraphics[width=7cm]{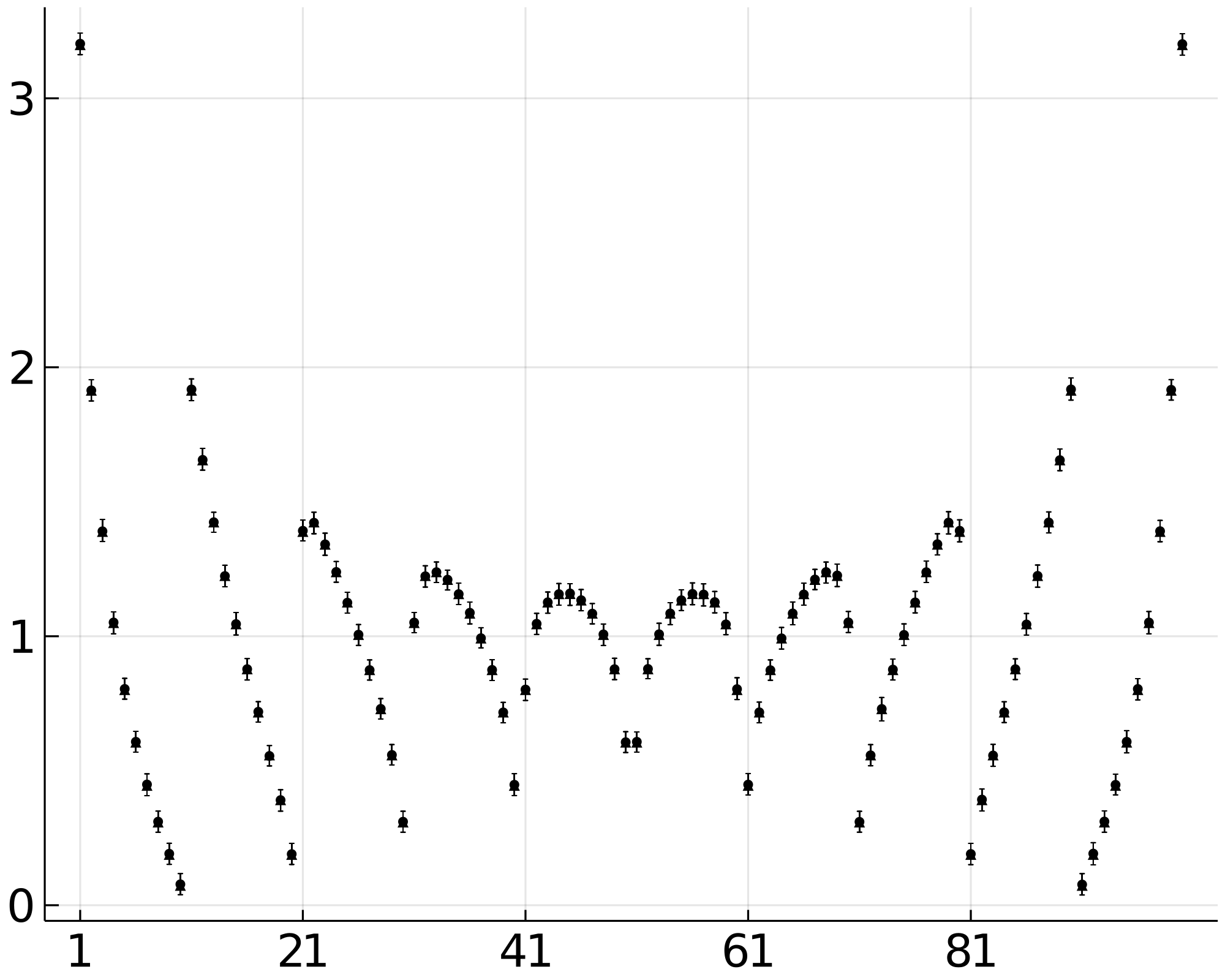}
}
\caption{\label{figure:prior1} Grid-uniform prior model - Prior mean (circle) and 95\% equal-tail credible interval (vertical line) of the copula density evaluated at 100 equidistant points of the bivariate sample space. The results are shown for different values of $\alpha^\star= \frac{\alpha}{|\nu^\rho|}$. In all cases, the centering copula model, $C_{0}$, is a bivariate Gaussian copula function with correlation equals to $0.5$. Panel (a), (b), (c), and (d), provides the results for $\alpha^\star=2$, $20$, $200$, and $2000$, respectively.  In each panel, the prior mode of the copula density is represented by triangles.}
\end{figure}

$C_{0,\rho}$ plays the role a centering parameter under the proposed prior and corresponds to the prior mode. On the other hand,  $\alpha$ plays the role of a precision parameter, since as $\alpha \rightarrow +\infty$, the prior variance $var(\pi(C|\rho,\alpha,C_0)) \rightarrow 0$. Figure~\ref{figure:prior1} illustrates the role of the parameters of the prior model. The figure displays the mean and credible interval of the prior distribution of the copula density at different points of the sample space.  

The impact of $\alpha$ scales with the size of the sets in the grid, meaning that for fine grids, $\alpha$ may have to be very large. To facilitate the prior elicitation process we consider the parameterization $\alpha^\star= \frac{\alpha}{|\nu^\rho|}$. For purposes of calculation, we note that using $C_{0,\rho}$ instead of $C_0$ as a reference copula produces the exact same prior, and calculating the prior using $C_{0,\rho}$ is much simpler, since its value is constant throughout each cell in the grid.

Assigning $\mathcal{D}$ to the squared-$L_2$ norm provides nicely interpretable parameters. However, it does not allow for the incorporation of prior information on the degree of smoothness of the copula function. For continuous copulas, it is often reasonable to expect that the value of a copula density at a point is similar to the value at nearby points. Let $\mathcal{V}^\rho$ be a set containing  the elements of $\nu^\rho$ in a given order. Let $\boldsymbol{W}$ be a symmetric matrix in which each entry $\boldsymbol{W}_{i,j}$ encodes information about the spatial relationship of the sets $\mathcal{V}^\rho_i$ and $\mathcal{V}^\rho_j$. Finally, let $\boldsymbol{D}_W$ be a diagonal matrix with $\boldsymbol{D}_{W_{i,i}}= \sum_{j=1}^{|\nu^\rho|} \boldsymbol{W}_{i,j}$. Borrowing ideas from models commonly used in spatial statistics and the nature of the grid-uniform model, we propose to take
$$
 \mathcal{D}\left(C,C_0 \right ) = \sum_{i=1}^{|\nu^\rho|}\sum_{j=1}^{|\nu^\rho|} \left( \boldsymbol{D}_W- \gamma \boldsymbol{W} \right)_{i,j} \int_{\mathcal{V}^\rho_i}(c(\boldsymbol{x})-c_0(\boldsymbol{x}) d\boldsymbol{x}   \int_{\mathcal{V}^\rho_j}(c(\boldsymbol{y}) - c_0(\boldsymbol{y}) )d \boldsymbol{y},
$$
where $\gamma >0$. Under this distance, the grid-uniform prior is given by
\begin{eqnarray}\nonumber  \label{prior2}
 \pi(C \mid \rho,\alpha,\boldsymbol{W}, \gamma, C_0) &\propto& \exp \left\{ - \frac{\alpha}{2}  \mathcal{D} \left(C, C_0 \right)  \right\} \times  I_{\mathcal{C}_{\rho}}(C),\\
  &=&  \exp\left\{ - \frac{\alpha}{2}  \overrightarrow{C-C_0}^T \left(\boldsymbol{D}_W- \gamma \boldsymbol{W} \right) \overrightarrow{C-C_0}\ \right\} \times   I_{\mathcal{C}_{\rho}}(C),
\end{eqnarray}
where $\overrightarrow{C-C_0}$ is the vector representation of $\left\{\int_Bc( \boldsymbol{x})-c_0(\boldsymbol{x})d \boldsymbol{x} : B\in\nu^\rho \right \}$ corresponding to the order induced by $\mathcal{V}^\rho$. Notice that the proposed prior corresponds to a truncated Gaussian conditional autoregressive (CAR) model, which allows for spatial correlation of nearby values, as specified by a smoothing parameter $\gamma >0$. It is worth noting that $\mathcal{D}$ is only a distance under certain conditions on $\boldsymbol{W}$ and $\gamma$ \citep[see, e.g.,][]{carbook}. 

A popular option is to set $\boldsymbol{W}$ such that $\boldsymbol{W}_{i,j}=\frac{1}{d_{ij}}$, where $d_{ij}$ is the distance between the centroids of $\mathcal{V}^\rho_i$ and $\mathcal{V}^\rho_j$ \citep{carprocs}. Another option is to set $\boldsymbol{W}$ as an adjacency matrix, where $\boldsymbol{W}_{i,j}=1$ if the sets $\mathcal{V}^\rho_i$ and $\mathcal{V}^\rho_j$ are grid-neighbors (in the usual intuitive sense), and  $\boldsymbol{W}_{i,j}=0$ otherwise. For practical purposes, the sparseness of the adjacency matrix produces a prior which is faster to compute, and  is a reasonable choice under most circumstances. When  $\boldsymbol{W}$ is the adjacency matrix, then $(\boldsymbol{D}_W)_{i,i}=| N_{\mathcal{V}^\rho_i}|$, where $N_B$ is the collection of grid-neighbors of the set $B$ and the distance reduces to the following expression
\begin{eqnarray}\nonumber
 \mathcal{D}(C,C_0)&=&\sum_{B\in \mathcal{V}^\rho}|N_B|\left(\int_B(c(\boldsymbol{x})-c_0(\boldsymbol{x}))d\boldsymbol{x}\right)^2-\\
   && \gamma\sum_{B\in \mathcal{V}^\rho} \left(\int_B(c(\boldsymbol{y})-c_0(\boldsymbol{y}))d\boldsymbol{y}\sum_{A\in N_B}\int_A(c(\boldsymbol{z})-c_0(\boldsymbol{z}))d\boldsymbol{z}\right).
\end{eqnarray}
Expressions for $\mathcal{D}$ as described above are not distances for all values of $\gamma$. To force  $\mathcal{D}$ to be a distance, we can pick $\gamma\in (\frac{1}{\lambda_1}, \frac{1}{\lambda_n})$, where $\lambda_1$ and $\lambda_n$ are the smallest and largest eigenvalues of $\boldsymbol{D}^{-1/2}_W W \boldsymbol{D}_W^{-1/2}$.  As discussed by \cite[see, e.g.,][section 6.4.3.3]{carbook}, the  spatial correlation is low unless $\gamma$ is close to 1. Because of this, a popular alternative is to consider $\gamma=1$, which is known as the Intrinsic CAR (ICAR) model. Under the ICAR, $\mathcal{D}$ is given by
$$
\mathcal{D}(C,C_0)=\sum_{i=1}^{|V^\rho|}\sum_{j=1}^{|\mathcal{V}^\rho|}W_{i,j}\left(\int_{\mathcal{V}^\rho_i} (c(\boldsymbol{x}))-c_0(\boldsymbol{x}))d\boldsymbol{x}-\int_{V_j^\rho}(c(\boldsymbol{y})-c_0(\boldsymbol{y}))d \boldsymbol{y}\right)^2.
$$
In addition, when $\boldsymbol{W}$ is the adjacency matrix, the distance reduces to
$$
\mathcal{D}(C,C_0)=\sum_{B\in\nu^\rho}\sum_{A\in N_B}\left(\int_B (c(\boldsymbol{x})-c_0(\boldsymbol{x}))d \boldsymbol{x}-\int_A (c(\boldsymbol{y})-c_0(\boldsymbol{y}))d\boldsymbol{y}\right)^2.
$$
In general, when $\gamma=1$ the latter expression is not a distance, but only to a pseudo-metric, since adding a constant to either $c$ or $c_0$ does not change the value of $\mathcal{D}(C,C_0)$. However, since $C$ and $C_0$ are both restricted to the space of $\rho$-uniform copulas, $\mathcal{D}$ does define a distance on the corresponding domain. Finally, note that all of the prior models share the algebraic structure of a truncated Gaussian distribution centered at $C_0^\rho$. In fact, all but the ICAR are exactly truncated Gaussian distributions. Therefore, the interpretation of $C_0$ and $\alpha$ remains intact.

\subsection{On the choice of hyper-parameters}

The prior depends on the choice of the grid $\rho$. The grid plays an equivalent role to the knots in the context of nonparametric regression based on splines. Rather than attempting to optimize the choice of a grid of reduced size and ``well'' located divisions, here we follow the approach proposed by \cite{eilers1996} in the context of penalized spline regression.
Specifically, we consider an equally spaced and fine grid, along a penalization induced by an ICAR model, given in expression (\ref{prior2}). The precise spacing of the grid can be chosen in relation to the available computational resources.
We have found that on mid-range modern hardware (a 3.9 GHz processor) a good posterior estimate for a  $50 \times 50$ grid (2401 free sets in the grid) can be computed in about 24 hours, whereas for a $10 \times10$ grid (81 free sets) a posterior estimate can be computed in about 2 minutes. This is due not only to the greater computational cost of calculating the prior and likelihood functions, but also since a larger number of MCMC iterations are required.

The parameters $\alpha^\star$  and $C_0$ have clear interpretations, and when prior information is available, it can be used to inform their choice. For situations when such information is not readily available, we propose suitable defaults. In this setting, selecting a single default $C_0$ around which to center the prior is difficult because, once specified, a single centering distribution may affect inference unduly. For instance, the use of the independent copula is highly informative because the lack of dependence is itself an extreme form of association structure. Rather than selecting a single centering copula function, one option is to consider a mixture of grid-uniform copula models by allowing the parameters of the centering copula function to be random. One possible choice  is to use the Gaussian copula family given by
$$
C_{0, \boldsymbol{R}}( \boldsymbol{x})=\Phi_{\boldsymbol{R}}(\Phi^{-1}(x_1), \Phi^{-1}(x_2), \ldots, \Phi^{-1}(x_d)),
$$
parametrized by the correlation matrix $\boldsymbol{R}$.

Choosing a prior for $\boldsymbol{R}$ is delicate since $\pi(C|\rho,\alpha,C_0)$ is known only up to a proportionality term, and this term depends on $C_0$ (and hence on $\boldsymbol{R}$). We can write out the full prior for $C$ as
$$
\pi(C|\alpha,\rho,\boldsymbol{R})=N(\boldsymbol{R}) \exp\left\{-\frac{1}{2}\alpha \times \mathcal{D}(C,C_{0,\boldsymbol{R}})\right\}
$$
where $N(\boldsymbol{R})$ is a normalizing constant.  By default, we consider a conditional prior for $\boldsymbol{R}$, such that 
$\pi(\boldsymbol{R} \mid \alpha, \rho, C_0)\propto \frac{1}{N(\boldsymbol{R})}$. This choice is computationally convenient for simulation, as described in Section \ref{hierarchicalmcmc}. It is difficult to find a closed form expression for this prior, but it is possible to characterize its behavior by means of simulation. Figure 3 illustrate the form of the prior in the bivariate case by considering $10 \times 10$ and $20 \times 20$ grids.
We observe that this prior does depend slightly on $\rho$ and $\alpha$, but the overall distribution is not greatly affected by the changing to a grid that has four times as many cells.
\begin{figure}
    \begin{centering}
    \subfigure[]{%
                \includegraphics[scale=0.1]{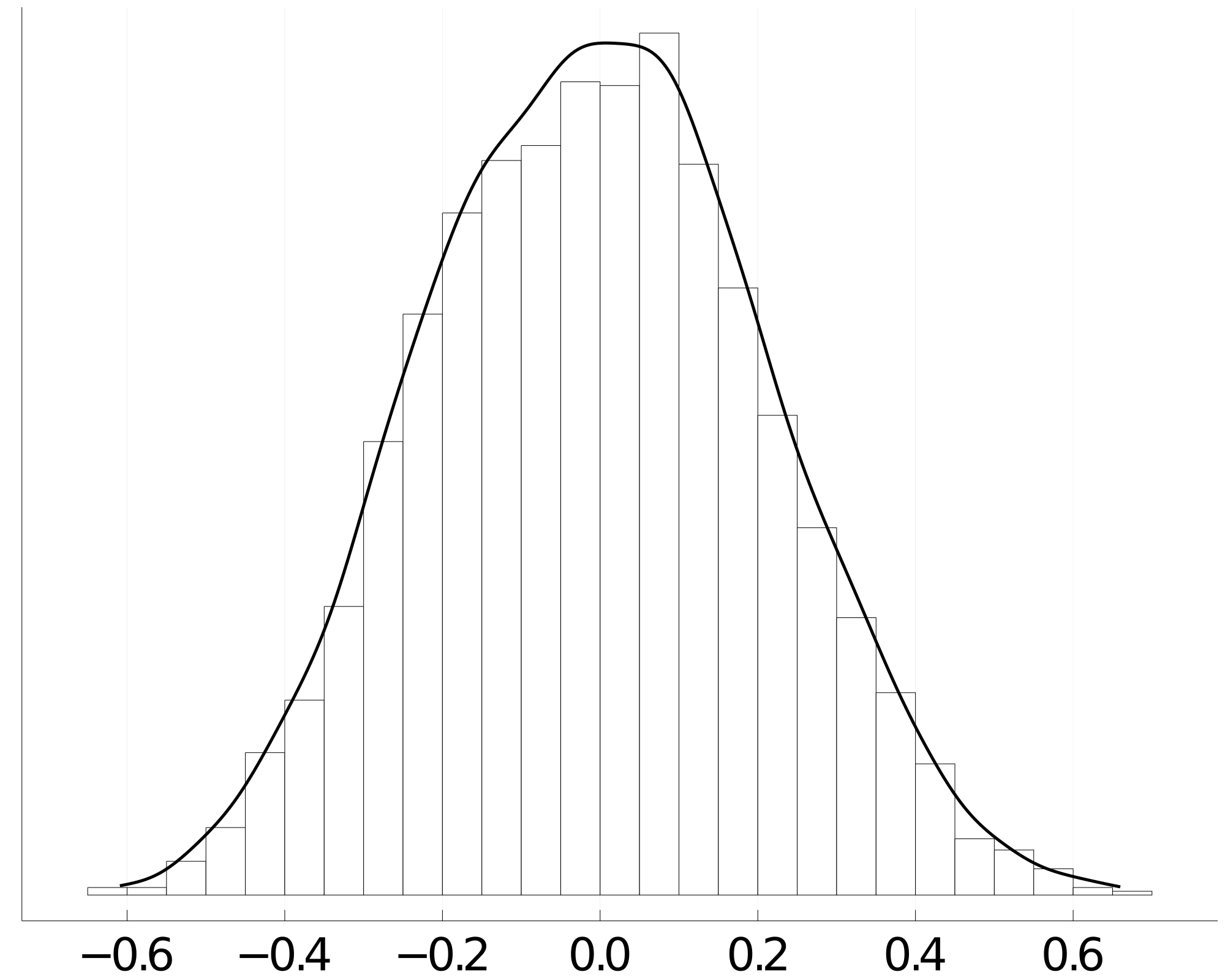}
    }
    \subfigure[]{%
              \includegraphics[scale=0.1]{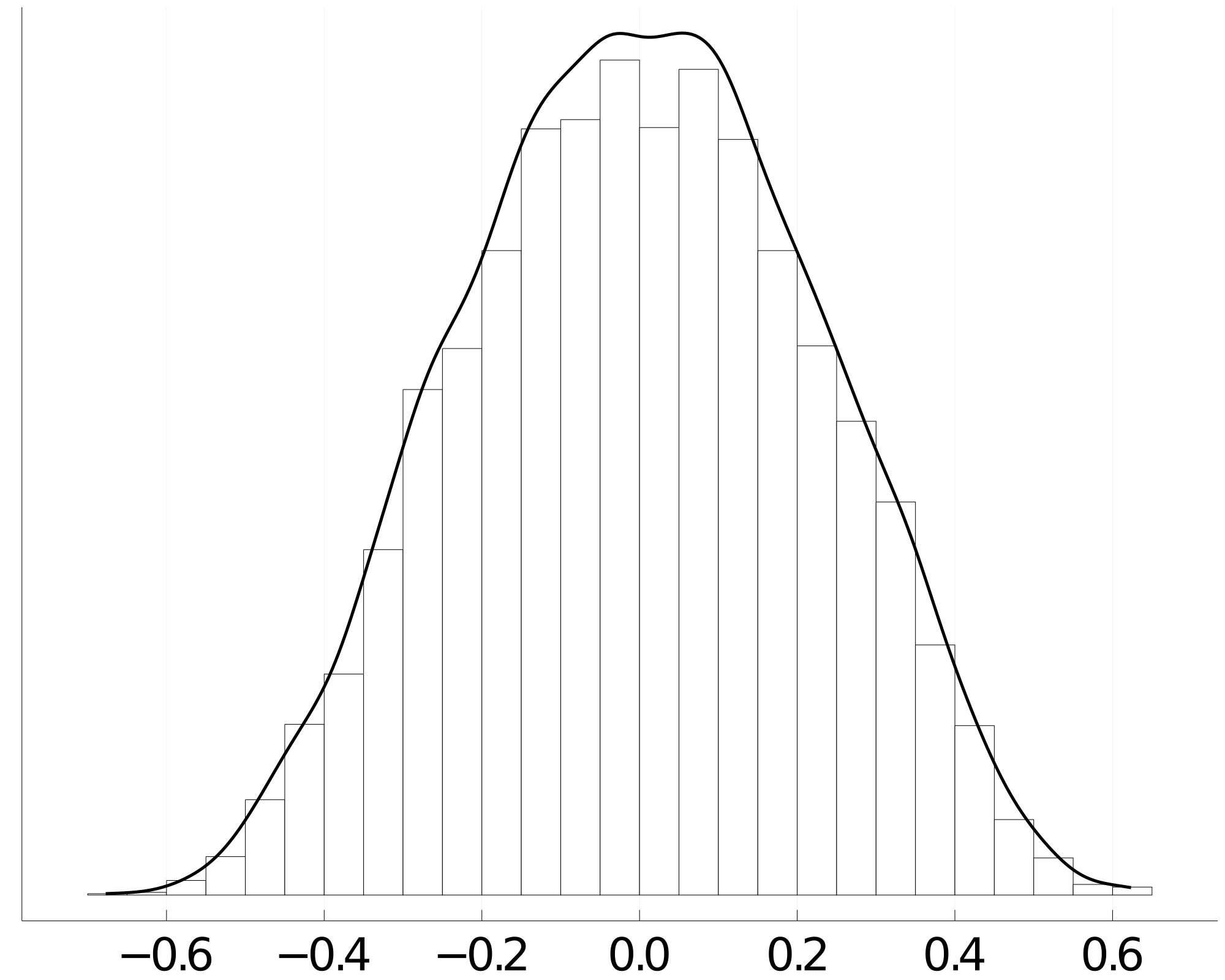}
     }\\         
     \subfigure[]{%
                \includegraphics[scale=0.1]{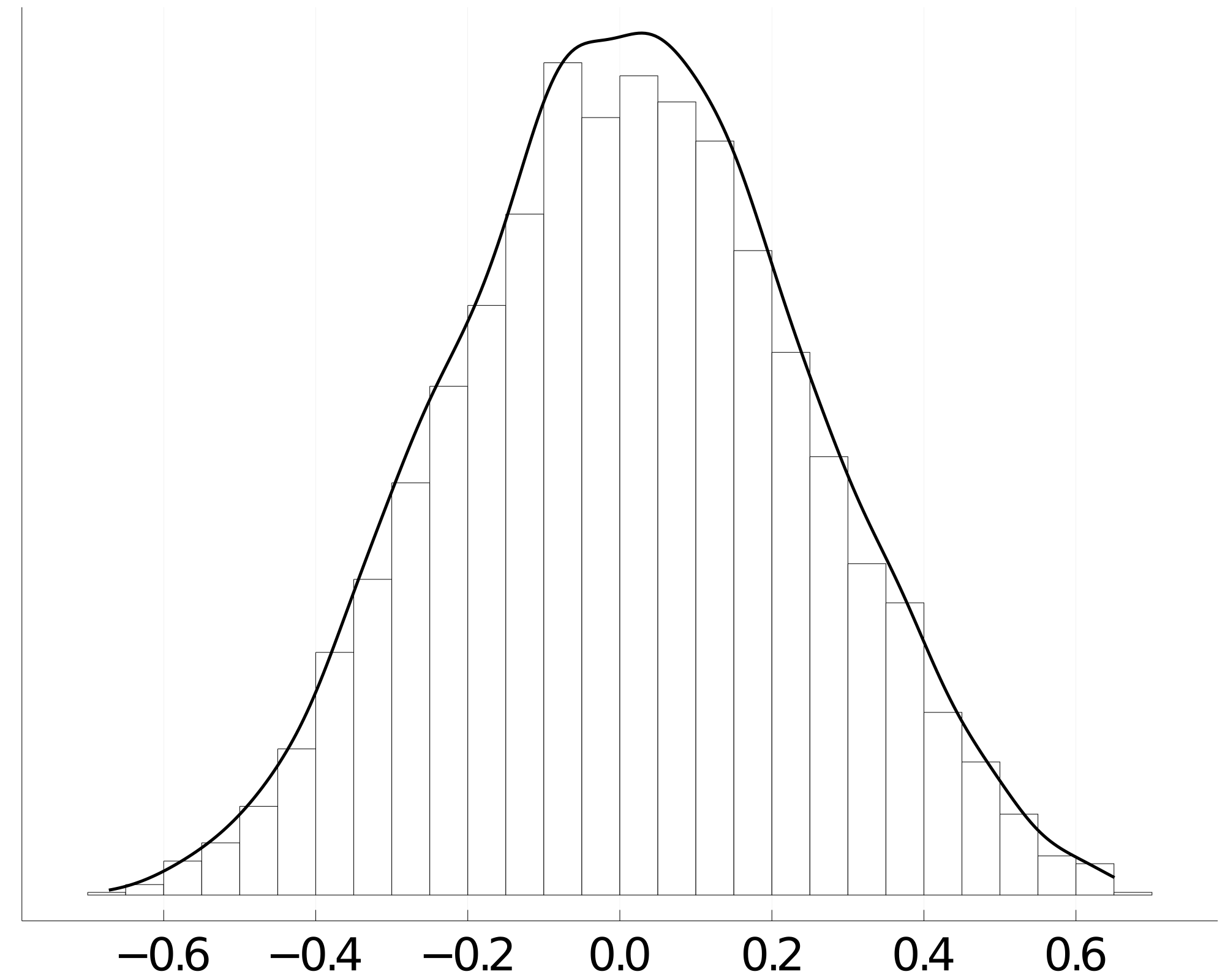}
     }
     \subfigure[]{%
                \includegraphics[scale=0.1]{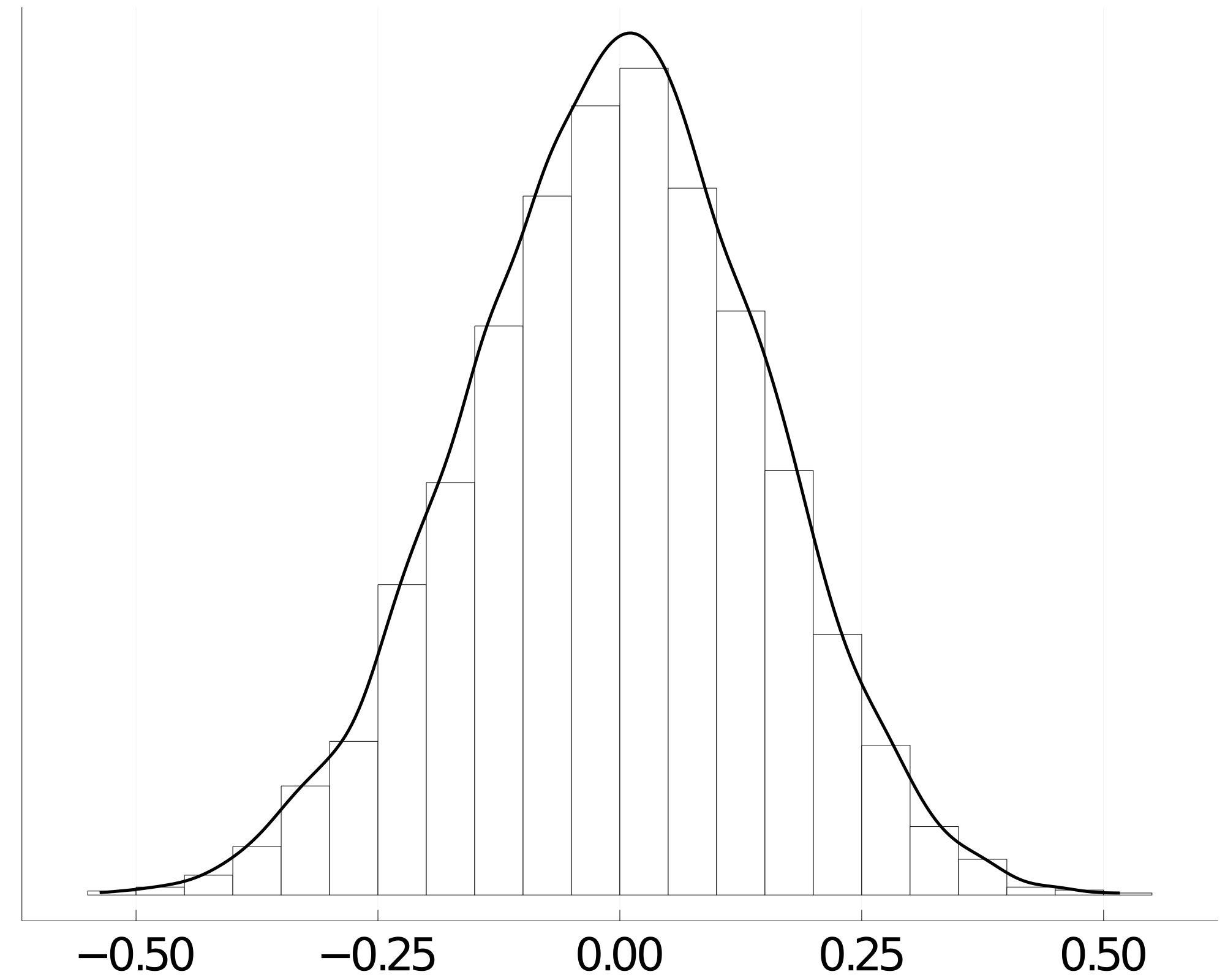}
     }\\
     \subfigure[]{%
            \includegraphics[scale=0.1]{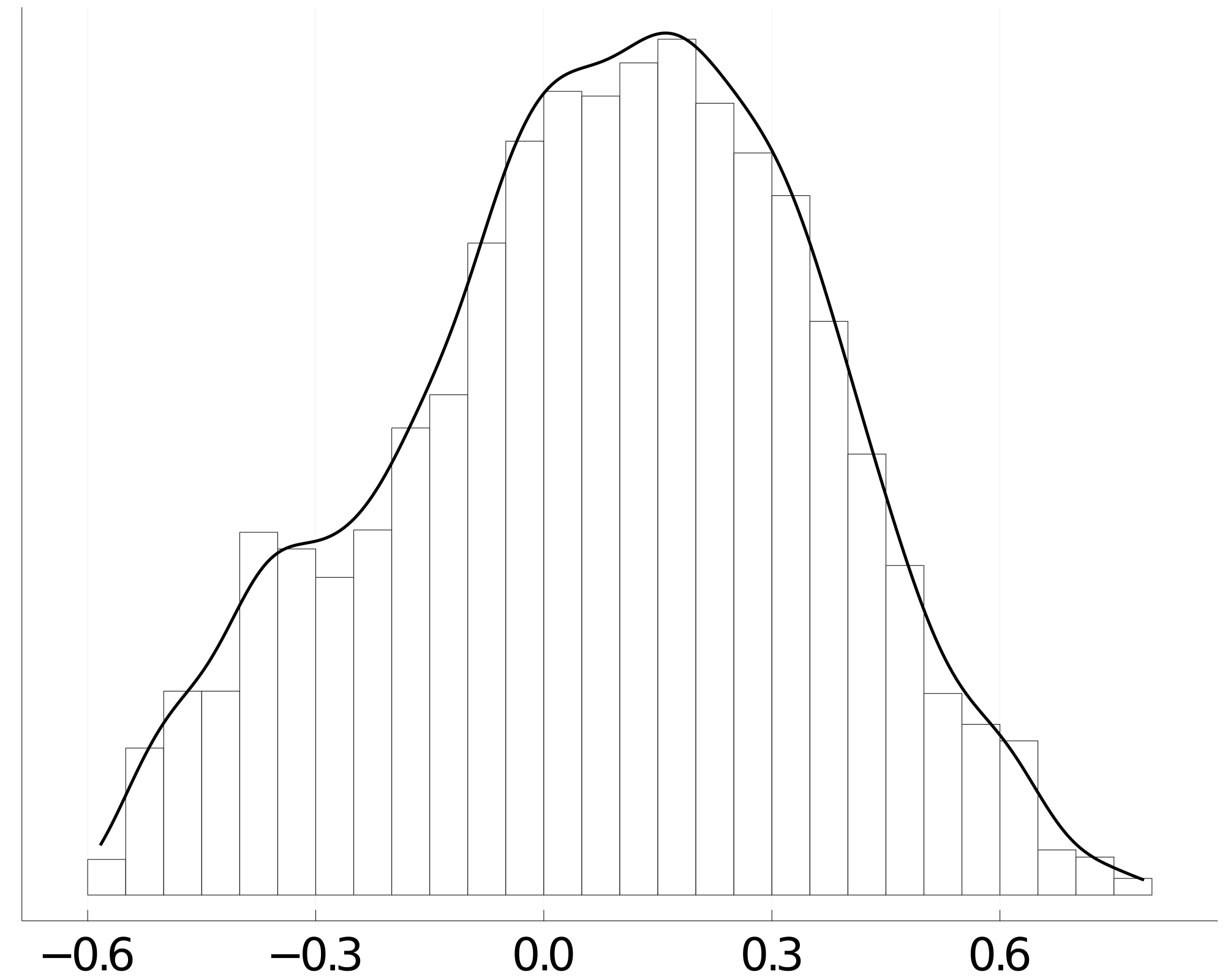}
     }       
     \subfigure[]{%
            \includegraphics[scale=0.1]{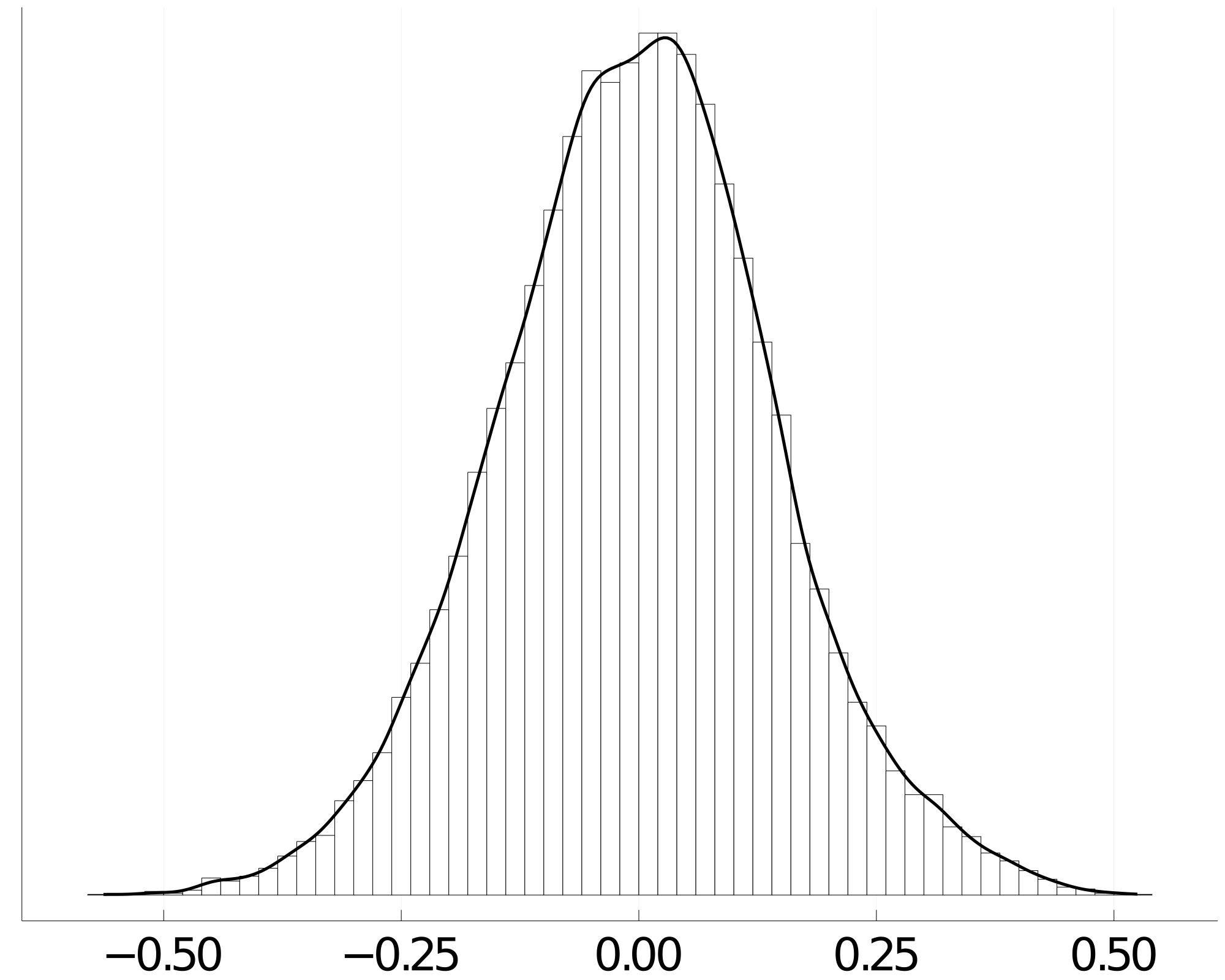}
      }      
        \caption{Prior distribution for the correlation coefficient of the centering Gaussian copula function. Panels (a), (c), and (e) display the results for an  evenly spaced $10 \times 10$ grid and with $\alpha^\star=25$, $100$, and $400$, respectively. Panels (b), (d), and (f) display the results for an  evenly spaced $20 \times 20$ grid and with $\alpha^\star=25$, $100$, and $400$, respectively.}
        \label{priorcorr}
    \end{centering}
\end{figure}
A similar procedure could be done to observe the prior for $\boldsymbol{R}$ in a higher dimensional case, but the results of this simulation of a prior over a high dimensional correlation matrix are quite difficult to interpret. Section \ref{hierarchicalmcmc} describes the algorithm to perform this simulation in the general case, but since the prior is difficult to understand, our implementation uses a fixed independent $C_0$ in dimensions higher than 2.

It is also possible to use a similar structure to allow $\alpha$ to be random, but the proportionality term for $\pi(C|\rho,\alpha,C_0)$ also depends on $\alpha$. A joint prior on $\alpha,\boldsymbol{R}$ with no closed form expression would be nearly impossible to interpret.

\subsection{An automatic MCMC algorithm}\label{mcmc}

We propose a Metropolis-within Gibbs algorithm for exploring the posterior distribution of the copula functions and the parameters associated with the marginal distributions.

\subsubsection{Updating $C$ using a random rectangle exchange as a proposal}

Rectangle exchanges provide us with a way to explore the space of $\rho$-uniform copulas. This movement can be used to generate proposals in the context of an Metropolis-Hastings (MH) algorithm. Let $C_{\rho}^{(b)}$ be the grid uniform copula which corresponds to the current state of the chain. Given,  $C_{\rho}^{(b)}$, we propose the candidate 
$\widetilde{C}_{\rho}$ using the following random rectangle exchange:
\begin{enumerate}

 \item[a)] Set $\widetilde{C}_{\rho} = C_{\rho}^{(b)}$ and pick $d_1$ and $d_2$ randomly from the set $\{1,\ldots,d\}$, and such that $d_1 < d_2$. The precise law of this selection does not matter so long as it is independent of  $C_{\rho}^{(b)}$ and every pair of coordinates has a positive probability of being selected. In practice, we will select them uniformly.
    \item[b)] Pick $a_{1}$ and $a_{2}$ from $\rho_{d_1}$ and pick $b_{1}$ and $b_{2}$ from $\rho_{d_2}$.
        Also, for all $k\in\{1,\ldots,d\}\setminus\{d_1,d_2\},$ pick $x_k\in\rho_k$, and set $
    \boldsymbol{p}_{(a_l,b_m)} = (x_1,\ldots,x_{i-1},a_l,,x_{i+1},\ldots,x_{j-1},b_m, x_{i+1}, \ldots,x_d)$, 
    where $l,m \in \{1,2\}$. The precise law of these selections does not matter so long as it is independent of  $C_{\rho}^{(b)}$ and every rectangle along the selected coordinates has positive probability of being selected. In practice, we will select them uniformly.
        
    \item[c)] Pick $\epsilon$ uniformly in the interval
        $$\left[\max\left\{-C\left(\nu^{\rho}_{\boldsymbol{p}_{(a_1,b_2)}} \right) ,-C\left(\nu^{\rho}_{\boldsymbol{p}_{(a_2,b_1)}} \right) \right\}, 
                \min\left\{ C\left(\nu^{\rho}_{\boldsymbol{p}_{(a_1,b_1)}} \right),C\left(\nu^{\rho}_{\boldsymbol{p}_{(a_2,b_2)}} \right) \right\}
        \right].$$
        
    \item[d)] Set \begin{eqnarray*}
                \widetilde{C}_{\rho}\left(\nu^{\rho}_{\boldsymbol{p}_{(a_1,b_1)}} \right) = C_{\rho}^{(b)}\left(\nu^{\rho}_{\boldsymbol{p}_{(a_1,b_1)}} \right)  - \epsilon, \\ 
            \widetilde{C}_{\rho}\left(\nu^{\rho}_{\boldsymbol{p}_{(a_1,b_2)}} \right) = C_{\rho}^{(b)}\left(\nu^{\rho}_{\boldsymbol{p}_{(a_1,b_2)}} \right) + \epsilon, \\
            \widetilde{C}_{\rho}\left(\nu^{\rho}_{\boldsymbol{p}_{(a_2,b_1)}} \right) = C_{\rho}^{(b)}\left(\nu^{\rho}_{\boldsymbol{p}_{(a_2,b_1)}} \right) +\epsilon,\\
            \text{and } \widetilde{C}_{\rho}\left(\nu^{\rho}_{\boldsymbol{p}_{(a_2,b_2)}}\right) =  C_{\rho}^{(b)}\left(\nu^{\rho}_{\boldsymbol{p}_{(a_2,b_2)}}\right)  - \epsilon.
    \end{eqnarray*}
\end{enumerate}

We denote by $q( \cdot \mid C_{\rho}^{(b)})$  to the candidate generating distribution induced by random rectangle exchange described by steps a) -- d). An interesting property of this candidate generating distribution is that it is symmetric, which simplifies the computation of the acceptance probability. This is explained by the uniform selection of  $\epsilon$ in the valid interval. The MH acceptance probability of the candidate $\widetilde{C}_{\rho}$ is given by
$$
 \max \left\{0, r\left(\widetilde{C_{\rho}},C^{(b)}_{\rho} \right)\right\}, 
$$
where
\begin{eqnarray}\nonumber
 \log\left(r \left(\widetilde{C_{\rho}},C^{(b)}_{\rho} \right) \right)&=& \log\left(\pi(\widetilde{C_{\rho}}|\rho,\alpha,C_0) \right) - \log \left(\pi(C_\rho^{(b)}|\rho,\alpha,C_0) \right) + \\\nonumber
    & & \ell(\widetilde{C_{\rho}},F_1,\ldots,F_d | y_1 \ldots y_n)-\ell(C_{\rho}^{(b)},F_1,\ldots,F_d | y_1 \ldots y_n),\\\nonumber
    &=& -\frac{\alpha}{2} \left( \mathcal{D}(\widetilde{C_{\rho}},C_0)-\mathcal{D}(C^{(b)}_{\rho},C_0) \right)+\\\nonumber
    & & \sum_{i=1}^n\sum_{j=1}^{|\nu^\rho|}\log \left( \frac {\widetilde{C_{\rho}}(B_j)-C^{(b)}_\rho(B_j)} {\lambda(B_j)} \right)
    I_{\left \{ \left(F_1(y_{i1}), \ldots, F_d(y_{id}) \right) \in B_j  \right \}}\left(\boldsymbol{y}_i\right),\\\nonumber
    &=& -\frac{\alpha}{2} \left( \mathcal{D}(\widetilde{C_{\rho}},C_0)-\mathcal{D}(C^{(b)}_{\rho},C_0) \right)-\\\nonumber
    & & \sum_{i=1}^n \sum_{k=1}^2 \sum_{l=1}^2 \log \left( \frac {\widetilde{C_{\rho}}\left(\nu^\rho_{\boldsymbol{p}_{(a_k,b_l)}} \right)-C^{(b)}_\rho \left(\nu^\rho_{\boldsymbol{p}_{(a_k,b_l)}} \right)} {\lambda \left(\nu^\rho_{\boldsymbol{p}_{(a_k,b_l)}} \right)} \right) \times \\\nonumber
    & & I_{\left \{ \left(F_1(y_{i1}), \ldots, F_d(y_{id}) \right) \in \nu^\rho_{\boldsymbol{p}_{(a_k,b_l)}}  \right \}}\left(\boldsymbol{y}_i\right).
\end{eqnarray}
Of note, if $\mathcal{D}$ is the squared-$L_2$ distance then the term $-\frac{\alpha}{2}(\mathcal{D}(\widetilde{C}_{\rho},C_0)-\mathcal{D}(C^{(b)}_\rho,C_0))$ can also be further simplified to:
$$
    - \frac{\alpha}{2} \left( \sum_{k=1}^2 \sum_{l=1}^2 \lambda(\nu^{\rho}_{\boldsymbol{p}_{(a_k,b_l)}}) \left((\widetilde{c_\rho}(\nu^{\rho}_{\boldsymbol{p}_{(a_k,b_l)}})- c_0(\nu^{\rho}_{\boldsymbol{p}_{(a_k,b_l)}}) )^2 - (c_\rho^{(b)}(\nu^{\rho}_{\boldsymbol{p}_{(a_k,b_l)}}) - c_0(\nu^{\rho}_{\boldsymbol{p}_{(a_k,b_l)}}) )^2\right)\right).
$$

In practice, we have found that this MH behaves well, with acceptance rates around 23\%. 

\subsubsection{Updating the marginal distributions}

There is no single technique which will work efficiently for the updating of the parameters  of the marginal distributions, and tuning the posterior sampler may be difficult. However, there are some algorithms which are effective for a broad scope of distributions, and that can do reasonably good posterior exploration without having to worry about tuning. A good starting point is the $t$-walk \citep{twalk}, which is a general purpose sampler for parametric continuous distributions. The $t$-walk is a MH algorithm which adapts to the scale of the target distribution and can sample well from most finite dimensional continuous distributions without tuning.

\subsubsection{Updating the centering copula hyper-parameter} \label{hierarchicalmcmc}

When working with a hierarchical prior that establishes a prior distribution for $C_0$ which depends on $\boldsymbol{R}$, updating $\boldsymbol{R}$ can be done by adding a kernel to the MCMC chain.
To update $C_0$, we use a variation of the metropolized hit-and-run algorithm \citep{hitandrun}, which makes proposals that are always valid correlation matrices.

In our specific case, we allow $\delta$ to be a pre-specified tuning parameter. We consider values between 0.3 and 1.0, as discussed by \cite {hitandrun}. To update $\boldsymbol{R}$, we propose a move from $\boldsymbol{R}^{(i)}$ to $\boldsymbol{R}^{(i+1)}=\boldsymbol{R}^{(i)}+\boldsymbol{H}$, by picking $\boldsymbol{H}$ as follows:
\begin{enumerate}
    \item[(1)]{Let $\xi^{(i)}$ be the least eigenvalue of $\boldsymbol{R}^{(i)}$}
    \item[(2)]{Pick a sequence of i.i.d. standard normal variables $z_{1,2}, z_{1,3}, \ldots z_{d-1,d}$.}
    \item[(3)]{Pick $\delta\sim N(0,\boldsymbol{r}^2)$ truncated to $(-\frac{\xi^{(i)}}{\sqrt{2}},\frac{\xi^{(i)}}{\sqrt{2}})$.}
    \item[(4)]{For $i<j$ set $h_{i,j}=\frac{\delta z_{i,j}}{\sum_{j=1}^{d-1}\sum_{l=j}^D z_{j,l}^2}$. Also set $h_{i,i}=0$ for all $i$, and for $i>j$ set $h_{i,j}=h_{j,i}$. Set the matrix $\boldsymbol{H}=[h_{i,j}]$}
\end{enumerate}
The acceptance probability is given by 
$$
\max\{0,r(\boldsymbol{R}^{(i+1)},\boldsymbol{R}^{(i)})\},
$$
where 
\begin{eqnarray}\nonumber
    \log(r(\boldsymbol{R}^{(i+1)},\boldsymbol{R}^{(i)}))&=& \log(\pi(C^{(i)}|\alpha^{(i)},\rho,\boldsymbol{R}^{(i+1)}))-\log(\pi(C^{(i)}| \alpha^{(i)}\rho, \boldsymbol{R}^{(i)})), \\\nonumber
    &=& \frac{1}{2}\alpha^{(i)}(\mathcal{D}(C,C_{0,R^{(i)}})-\mathcal{D}(C, C_{0,R^{(i+1)}})).
\end{eqnarray}

Our prior is selected so that $\log(N(\boldsymbol{R}))-\log(N(\boldsymbol{R}))$ cancels and we are not hampered by our inability to calculate it.

Note that for a two dimensional copula, $\boldsymbol{R}$ depends only the single dimensional correlation coefficient, $\boldsymbol{r}$
and the hit and run algorithm reduces to a standard Random Walk Metropolis kernel \citep{mcmc}.

\section{Illustrations}\label{illustrations}

We illustrate the behavior of the proposed model by means of the analysis of simulated data. Functions implementing the MCMC algorithms employed in these analyses were written in Julia and are available upon request to the authors.

\subsection{Estimation of parametric and non-standard copula functions}

\begin{figure}
    \begin{centering}
       \subfigure[$N=500$]{%
          \includegraphics[scale=0.1]{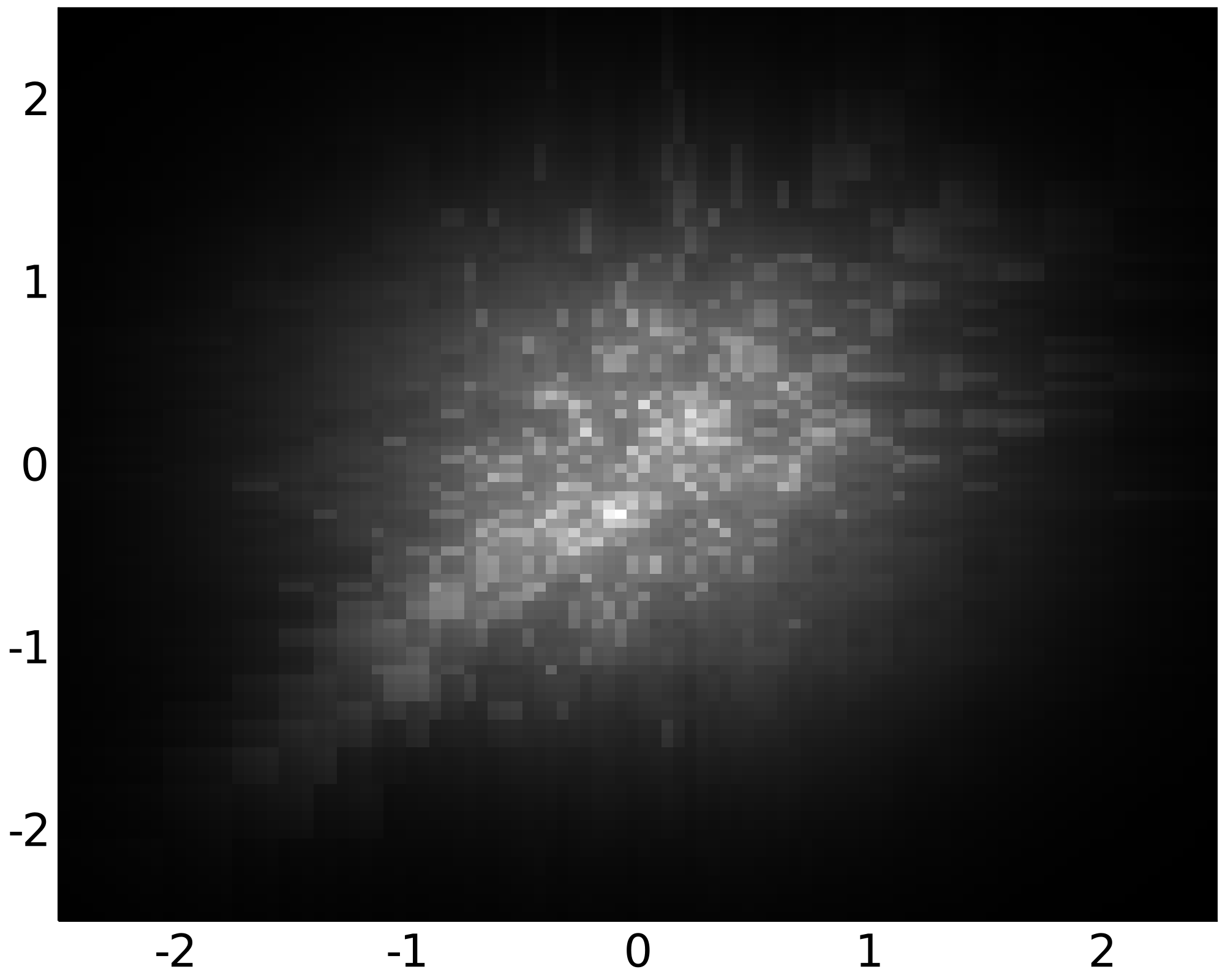}
         }
       \subfigure[$N=1000$]{%
          \includegraphics[scale=0.1]{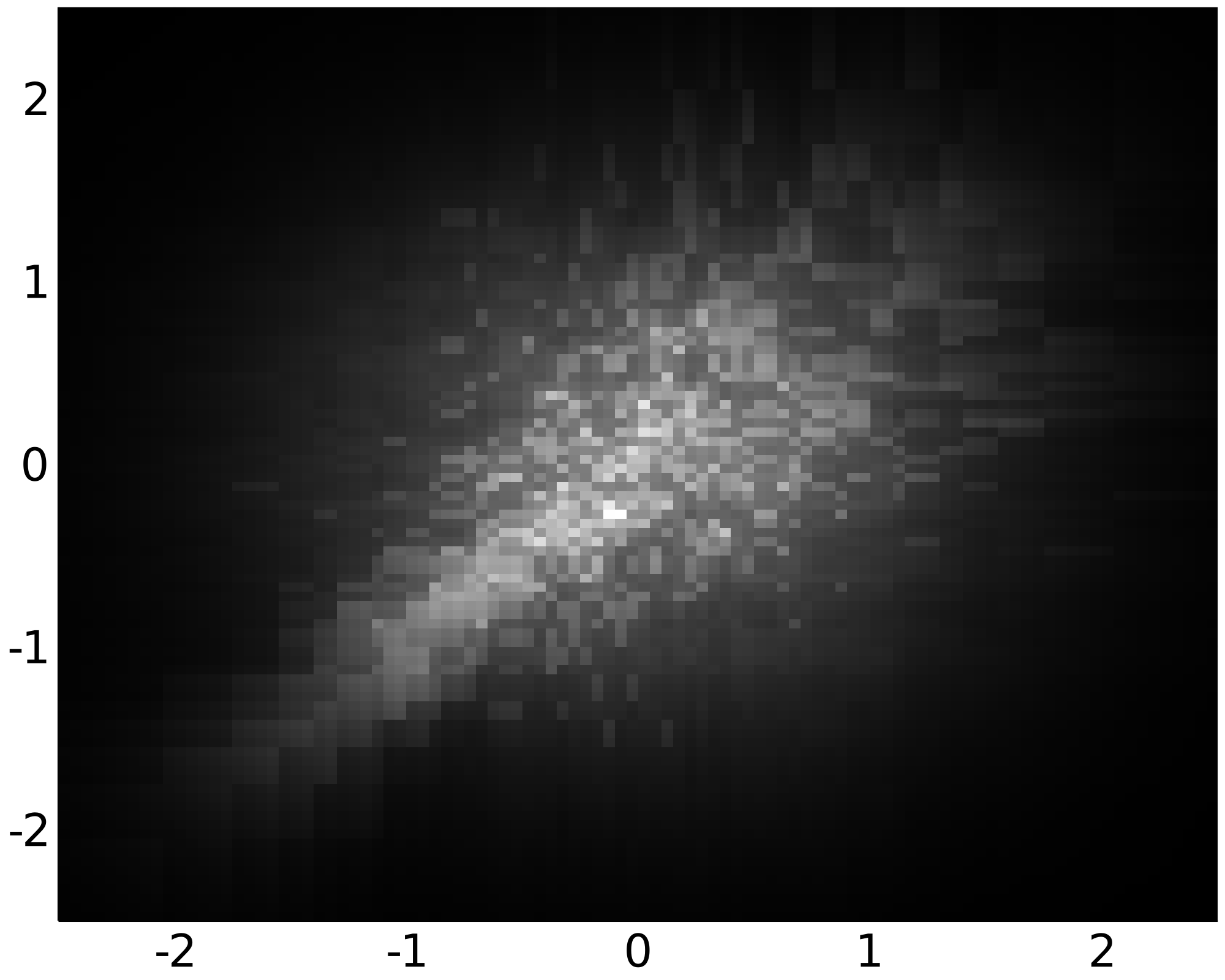}
         }\\
         \subfigure[$N=5000$]{%
          \includegraphics[scale=0.1]{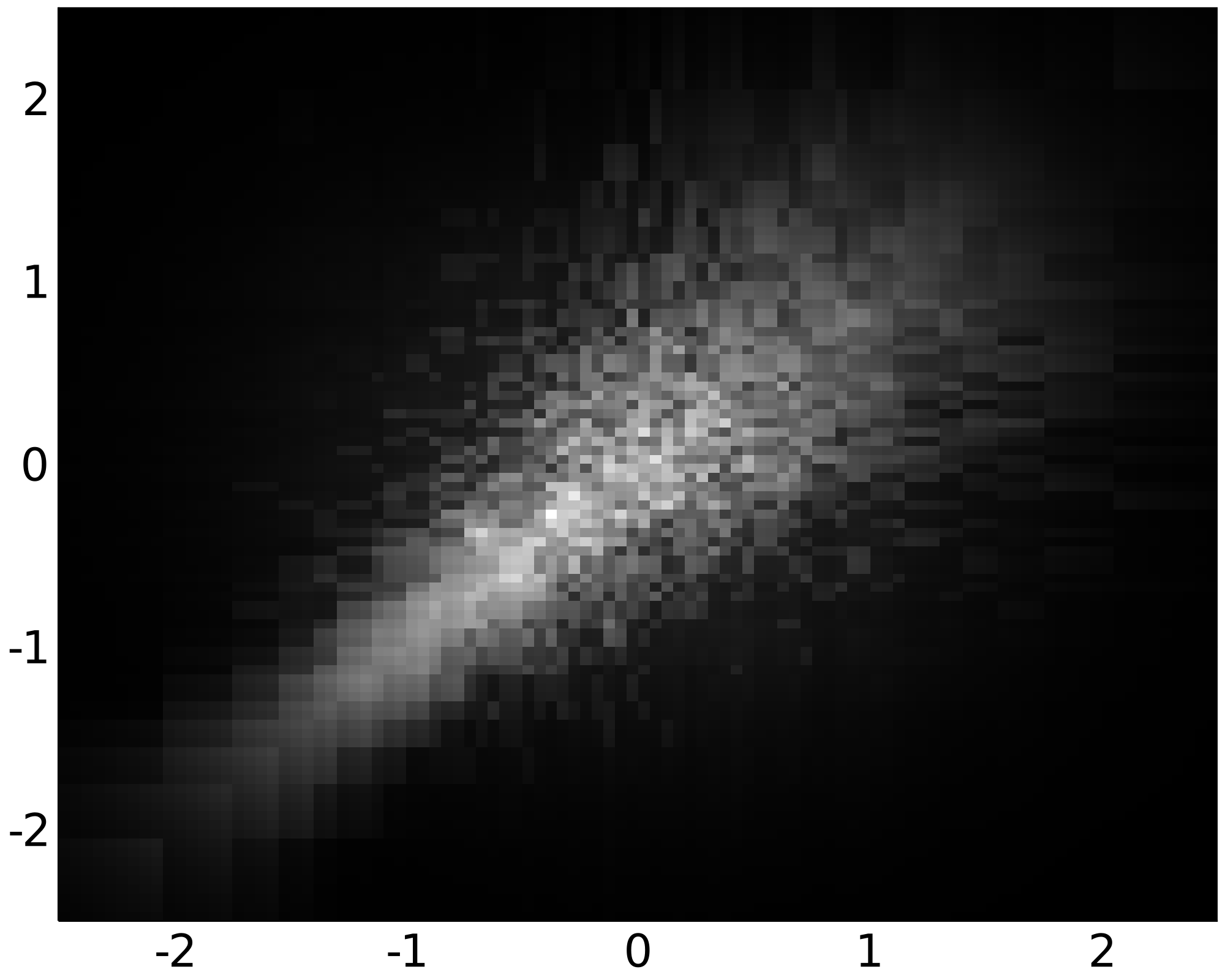}
         }
         \subfigure[$N=10000$]{%
          \includegraphics[scale=0.1]{jerarquicoclayton5000alpha1000000.png}
         }\\
         \subfigure[True model]{%
          \includegraphics[scale=0.1]{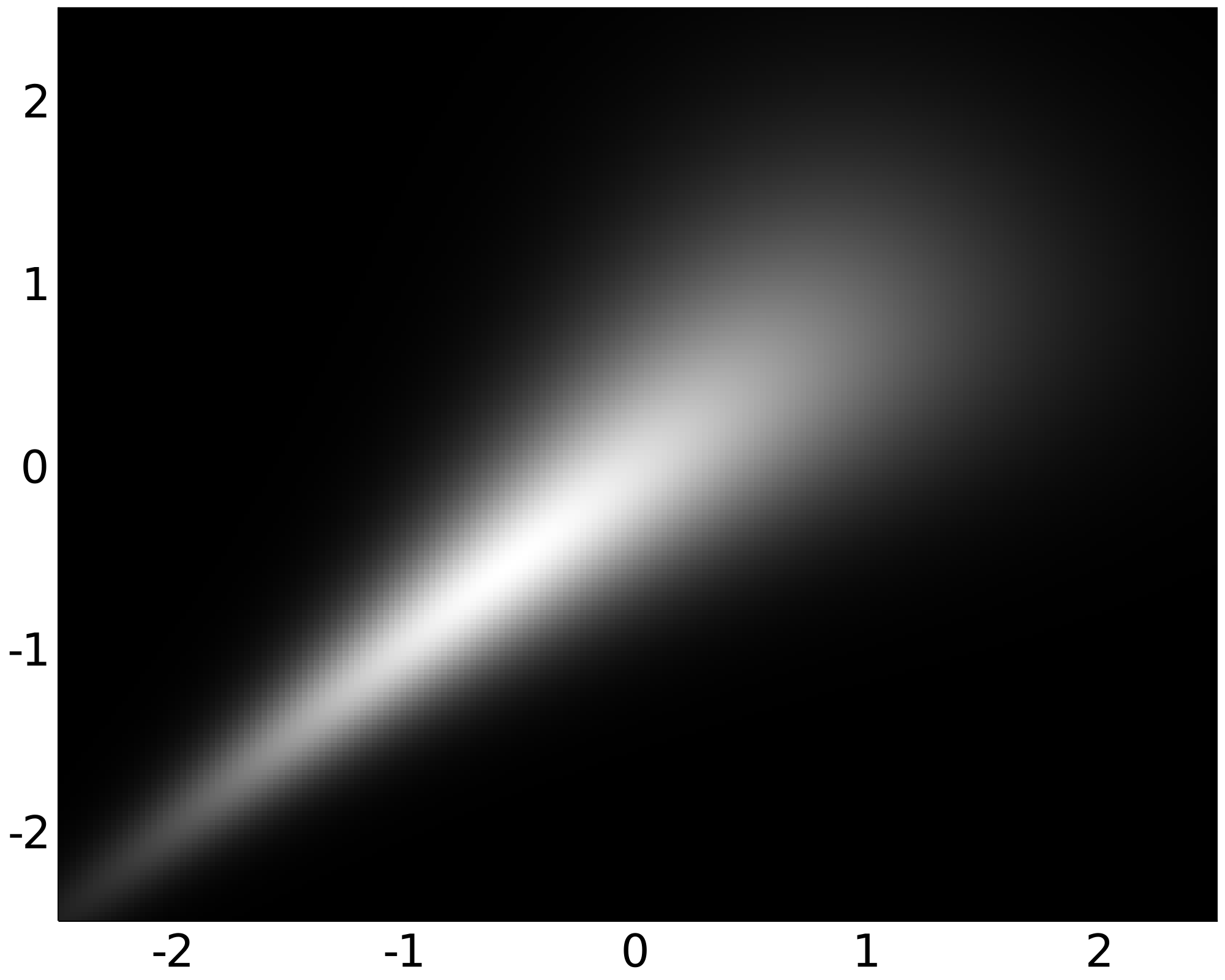}
         } 

        \caption{Model 1 (Clayton copula). Posterior mean of the bivariate density function. Panel (a) - (d) show the results for $N=500$, $1000$,
        $5000$, and $10000$, respectively. Panel (e) displays the true model.}
        \label{resultadosclayton}
    \end{centering}
\end{figure}

\begin{figure}
    \begin{centering}
           \subfigure[$N=500$]{%
          \includegraphics[scale=0.1]{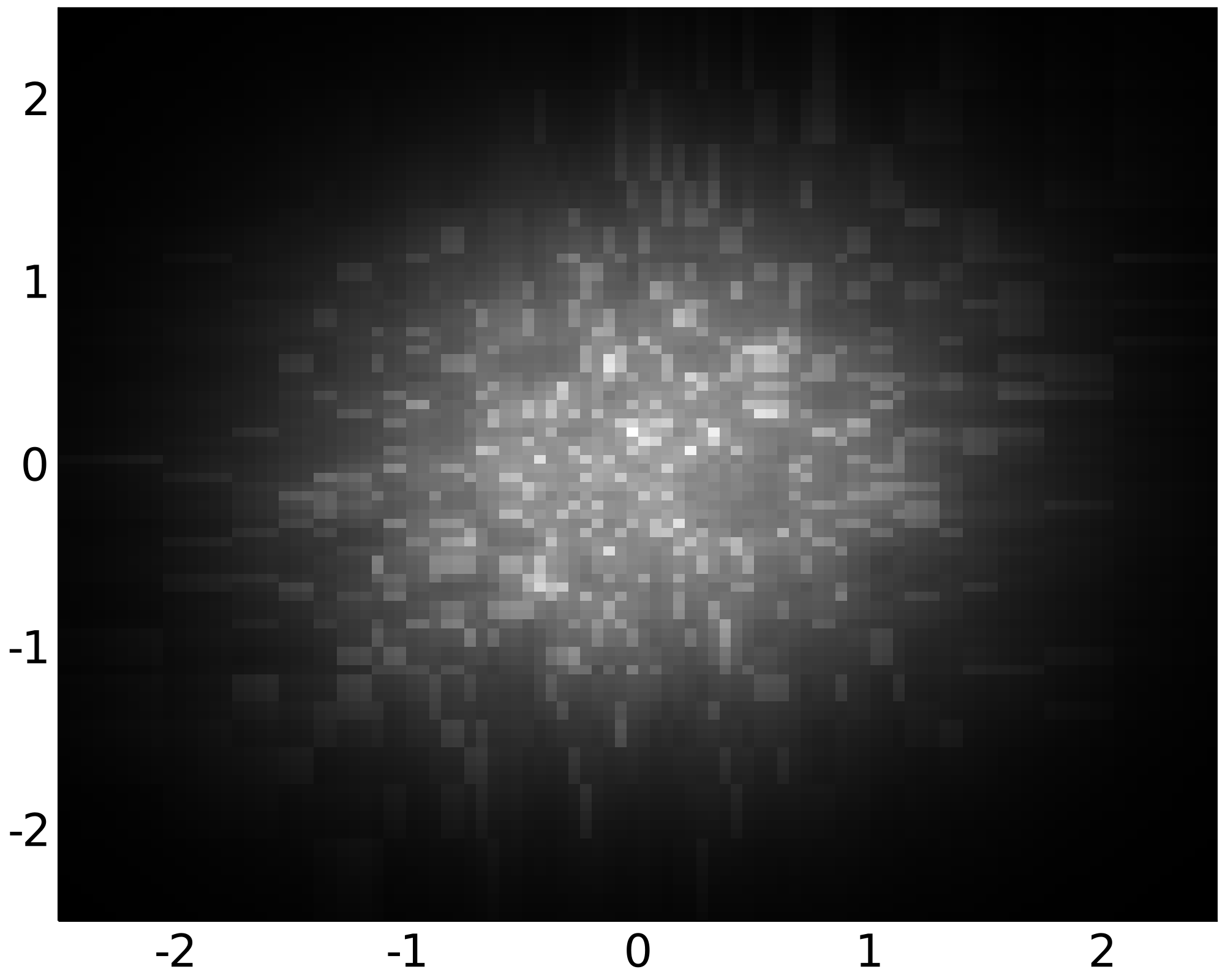}
         }
       \subfigure[$N=1000$]{%
          \includegraphics[scale=0.1]{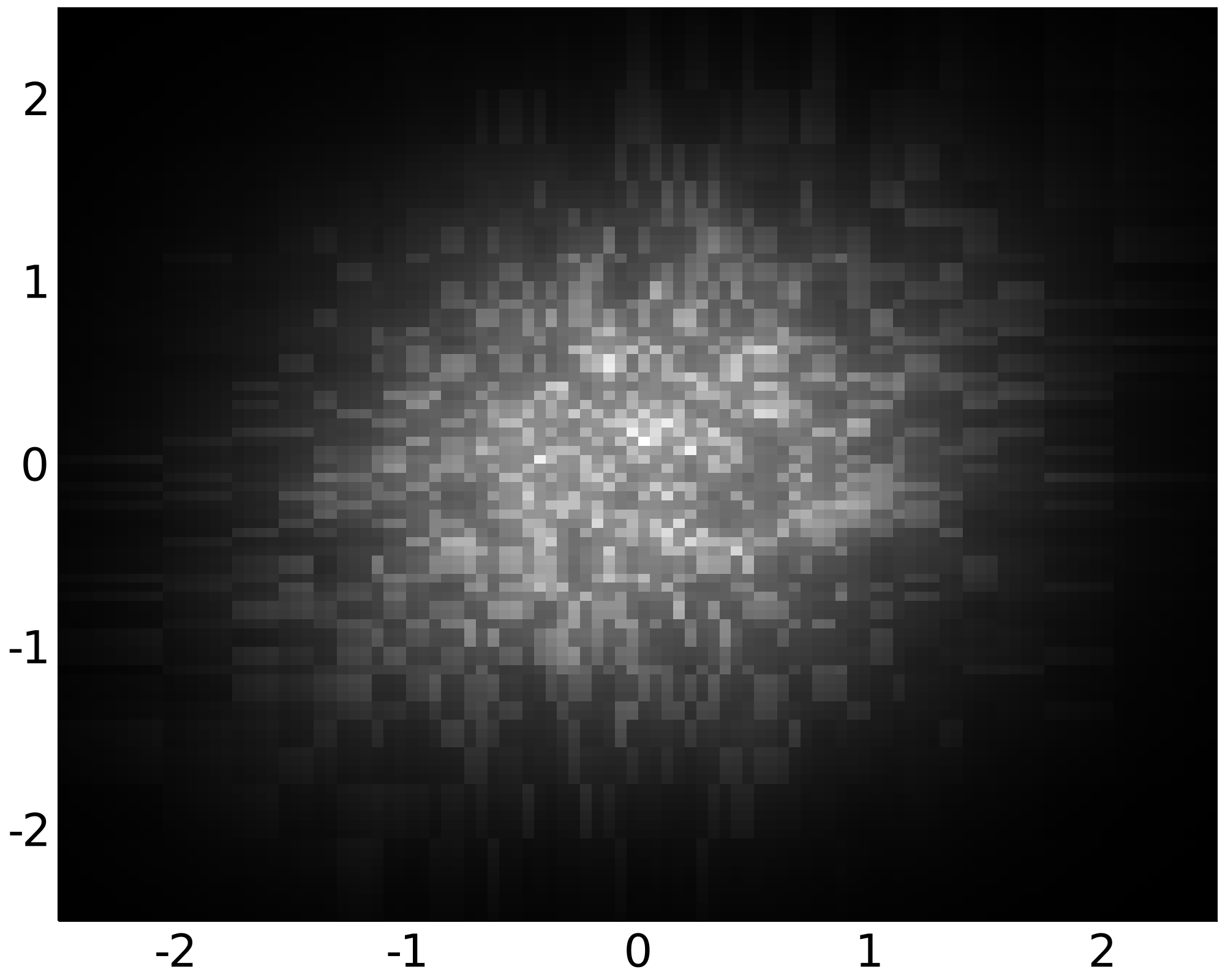}
         }\\
         \subfigure[$N=5000$]{%
          \includegraphics[scale=0.1]{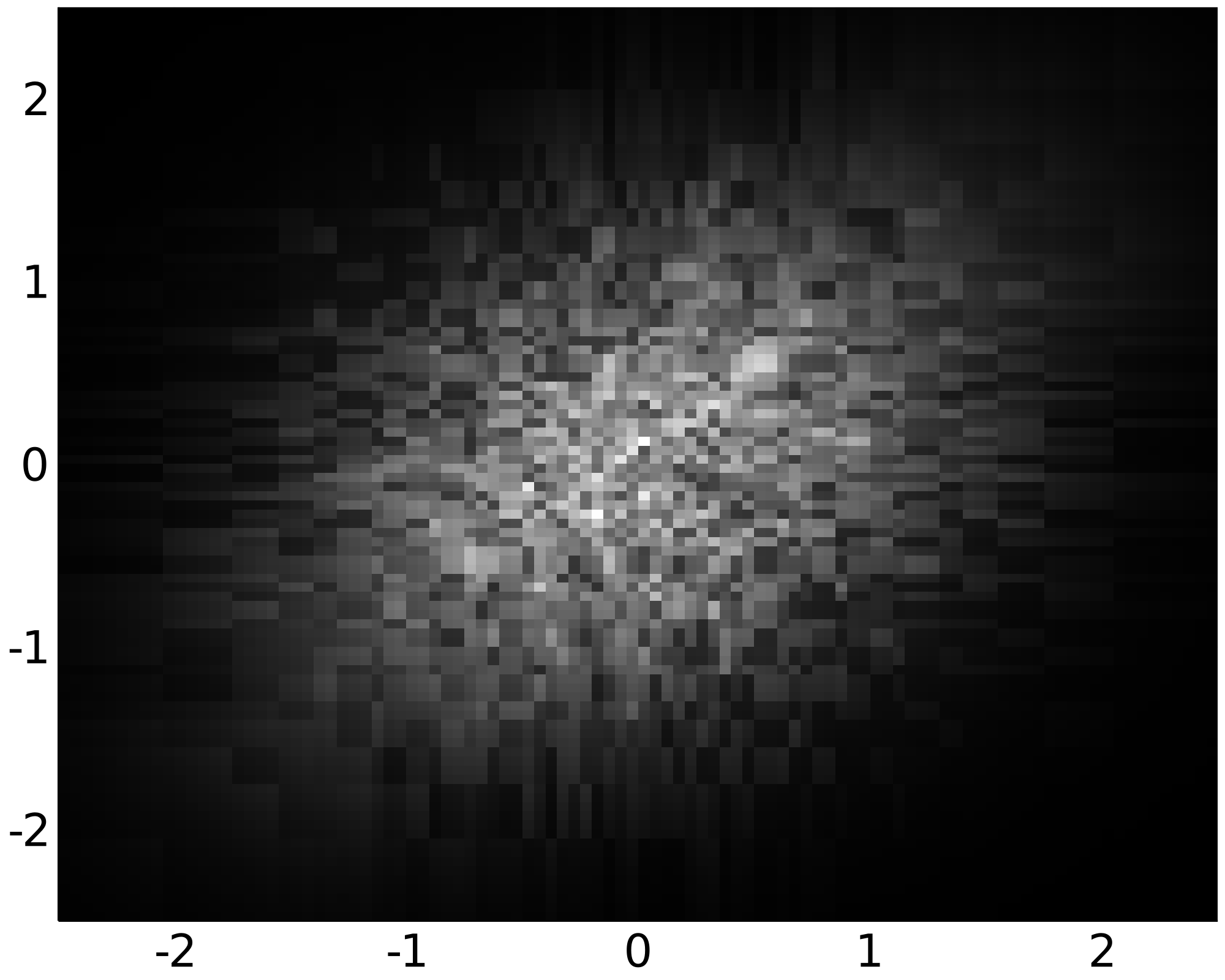}
         }
         \subfigure[$N=10000$]{%
          \includegraphics[scale=0.1]{jerarquicogauss5000alpha1000000.png}
         }\\
         \subfigure[True model]{%
          \includegraphics[scale=0.1]{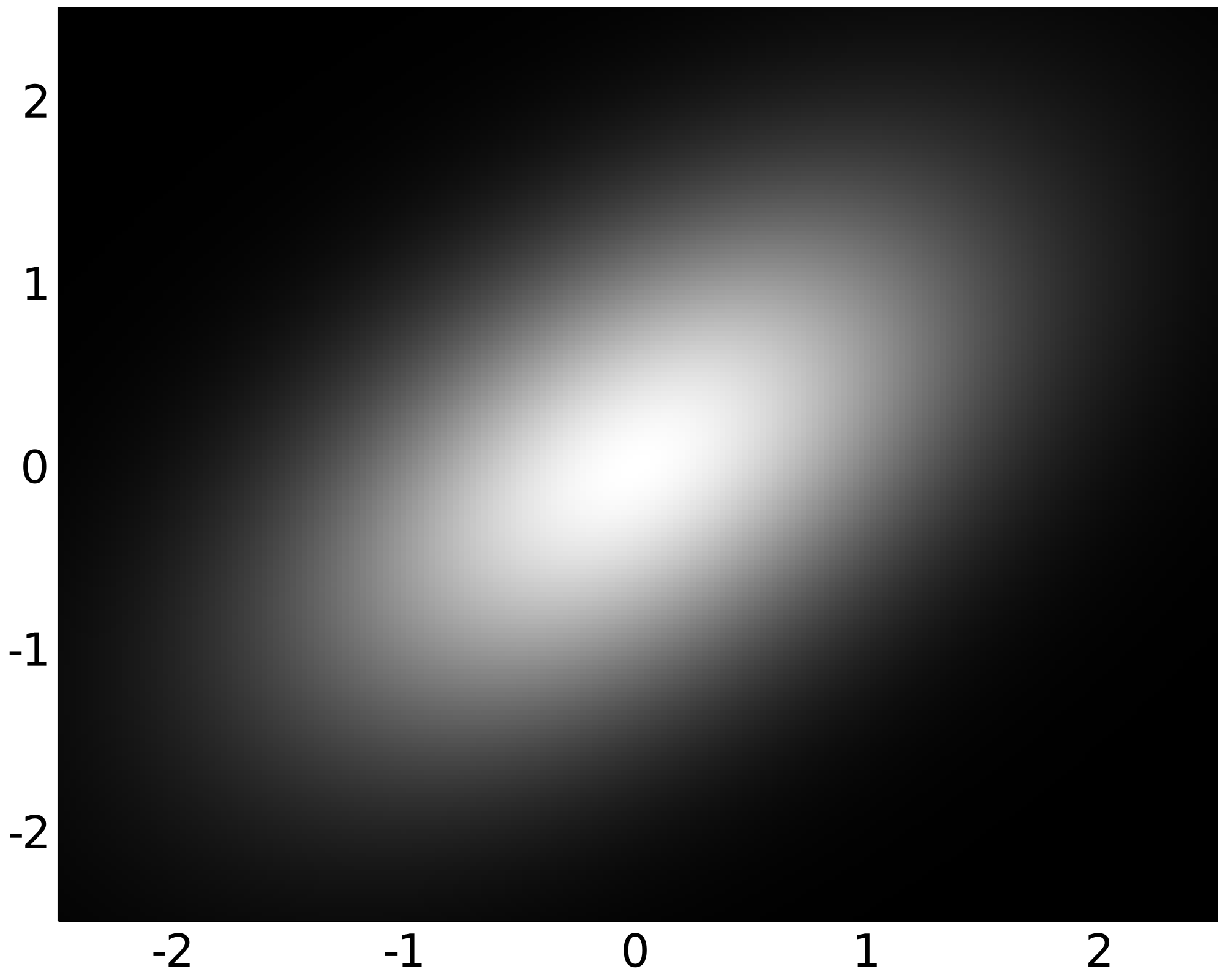}
         } 
        \caption{Model 2 (Gaussian copula). Posterior mean of the bivariate density function. Panel (a) - (d) show the results for $N=500$, $1000$, $5000$, and $10000$, respectively. Panel (e) displays the true model.}
        \label{resultadosgaus}
    \end{centering}
\end{figure}

\begin{figure}
    \begin{centering}
           \subfigure[$N=500$]{%
           
          \includegraphics[scale=0.1]{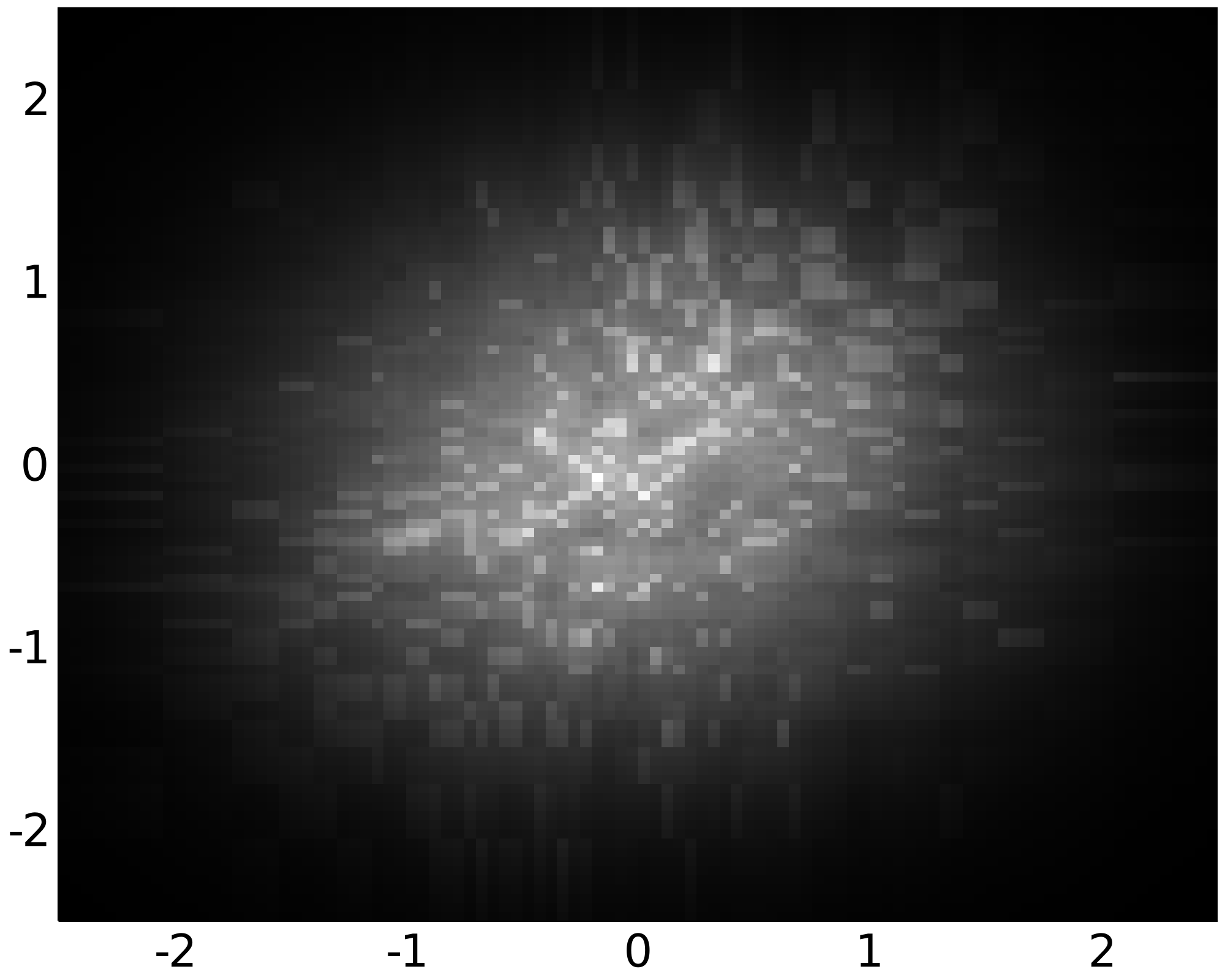}
         }
       \subfigure[$N=1000$]{%
          \includegraphics[scale=0.1]{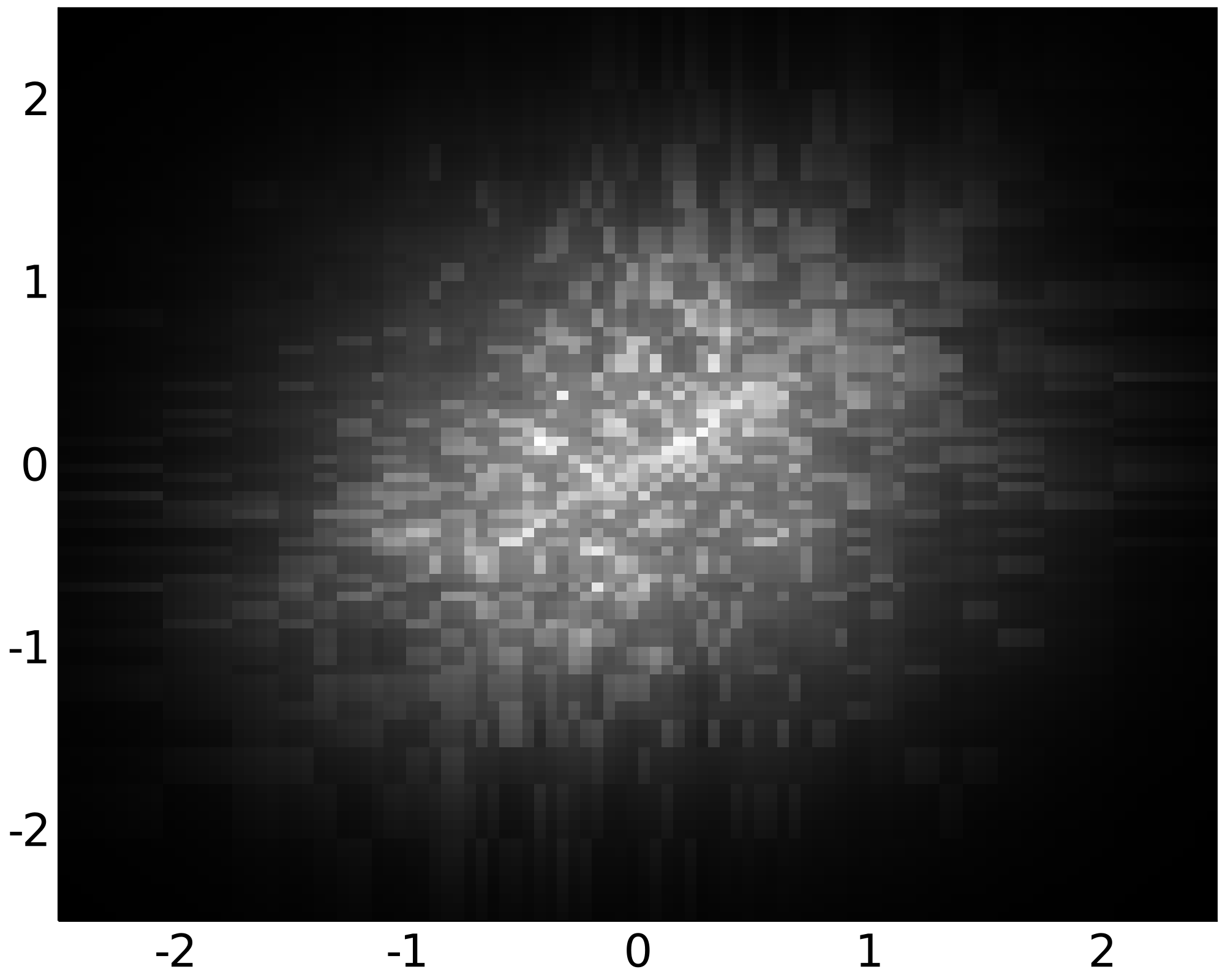}
         }\\
         \subfigure[$N=5000$]{%
          \includegraphics[scale=0.1]{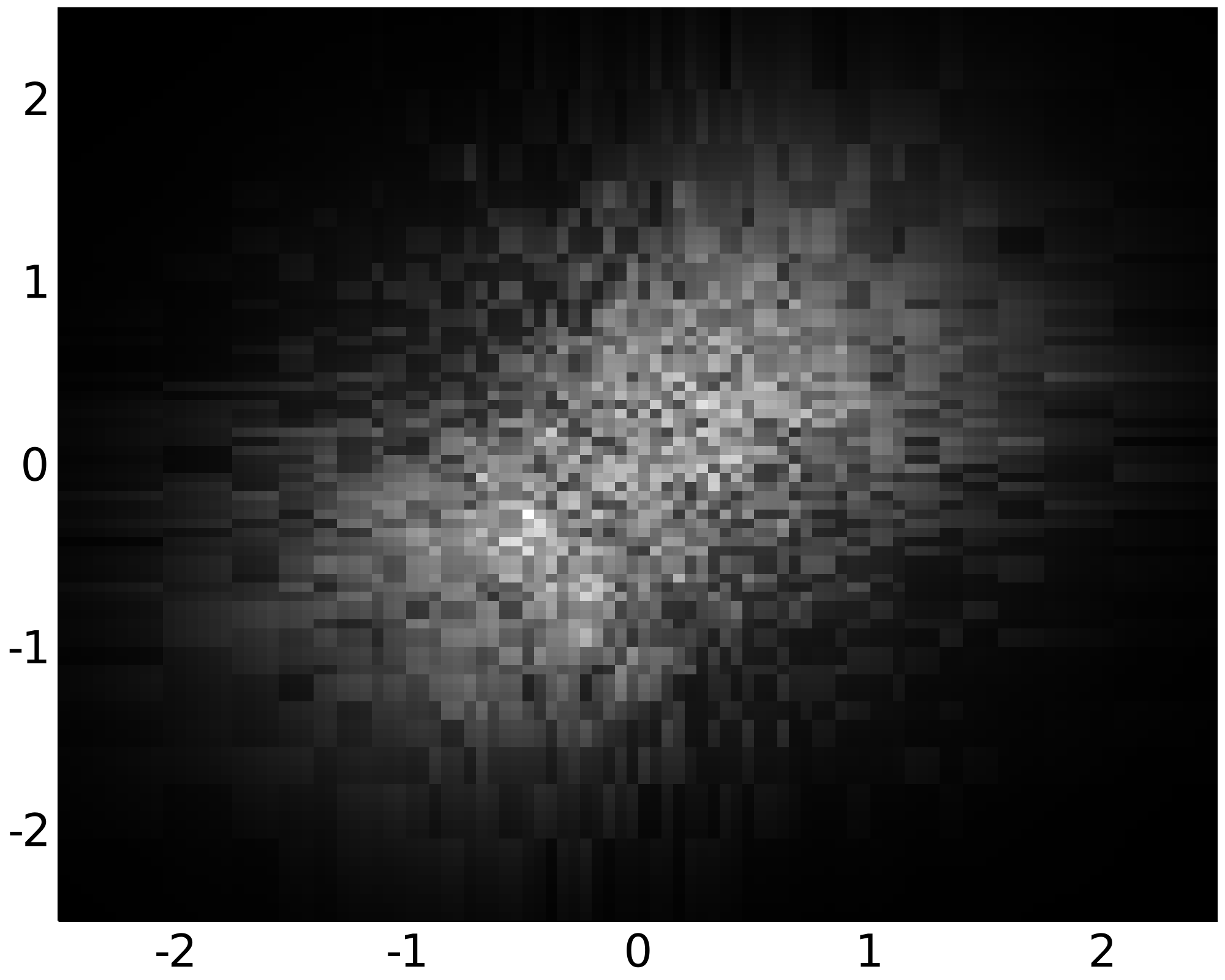}
         }
         \subfigure[$N=10000$]{%
          \includegraphics[scale=0.1]{jerarquicomezcla5000alpha1000000.png}
         }\\
         \subfigure[True model]{%
          \includegraphics[scale=0.1]{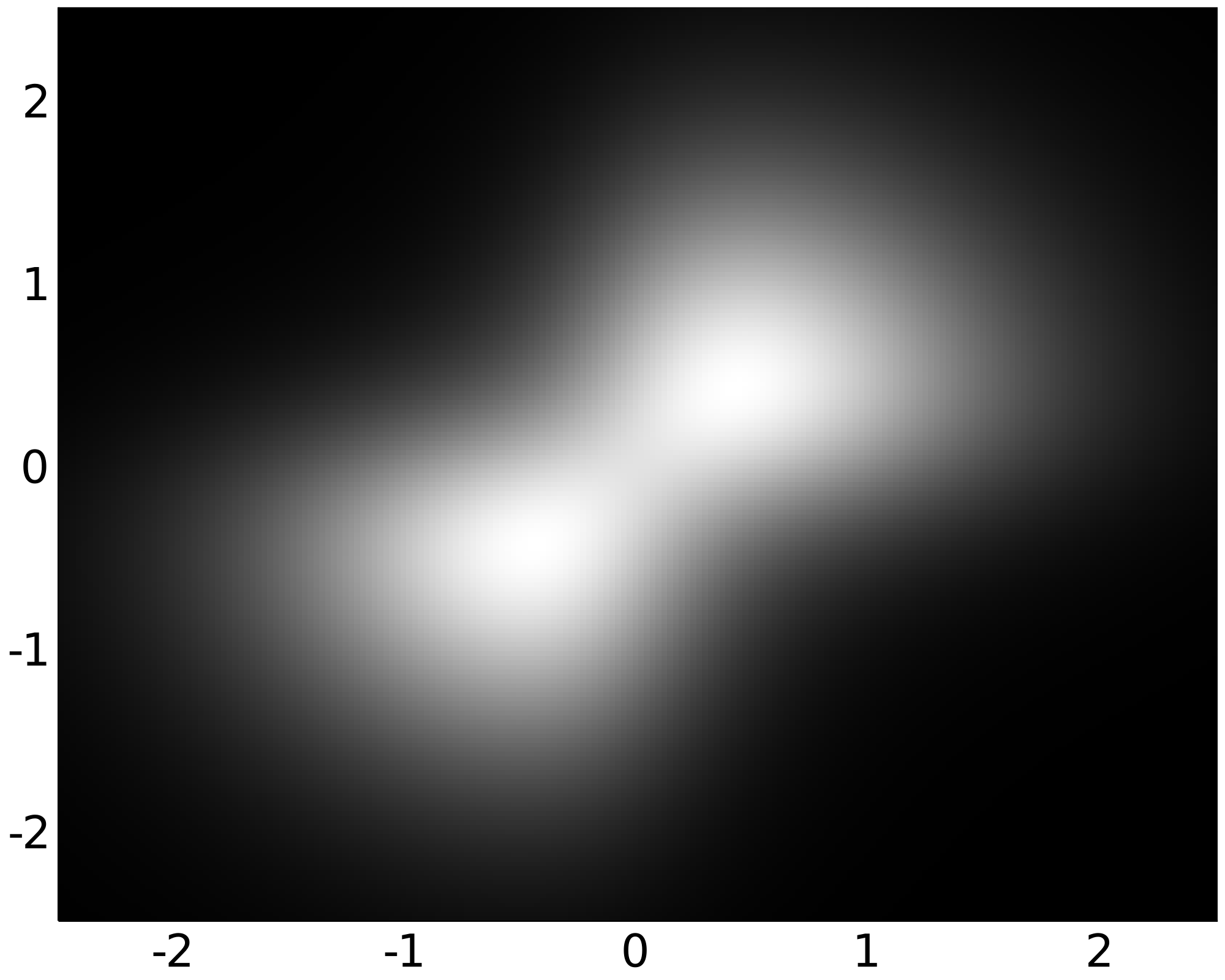}
         } 
        \caption{Model 3 (Copula of a Gaussian mixture model). Posterior mean of the bivariate density function. Panel (a) - (d) show the results for $N=500$, $1000$, $5000$, and $10000$, respectively. Panel (e) displays the true model.}
        \label{resultadosmezcla}
    \end{centering}
\end{figure}

To illustrate that the proposal model does not overfit the data when a parametric copula model holds and that is able to capture deviations from the standard parametric models with finite sample sizes, we consider bivariate models with Gaussian $(0,1)$ marginals, under the following copula functions:
\begin{itemize}
 \item {\bf Model 1:} A Clayton copula with parameter $\theta=3$, given by $C_\theta (x_1,x_2)=(\max\{u^{-\theta}+v^{-\theta}-1; 0\})^{-1/\theta}$.
\item {\bf Model 2:} A Gaussian copula with correlation 0.5.
\item {\bf Model 3:} A copula arising from a two component mixture of Gaussian distributions, both with identity covariance matrix,  and centered at  $(1,1)$ and $(-1,-1)$, respectively.
  \end{itemize} 
Figures~\ref{resultadosclayton}, \ref{resultadosgaus}, and \ref{resultadosmezcla} display the true models under consideration. For each model, we simulate a single data set of size $N=500$, $1000$, $5000$ and $10000$. For each simulated dataset we fit our proposed model by considering a $50 \times 50$ grid, the hierarchically centered prior, with the ICAR correlation structure described in Section \ref{priors}, and $\alpha^\star=400$. In these analyses we assume the marginals distributions to be known. We create a Markov chain of (conservative) of size 2,000,000 using the automatic algorithm described in Section \ref{mcmc}. We consider a burn-in period of 20,000 and a thinning of 1,000. Figures \ref{resultadosclayton}, \ref{resultadosgaus}, and \ref{resultadosmezcla} show the posterior mean under the different models and sample sizes. 

\begin{figure}
    \begin{centering}
     \subfigure[]{%
            \includegraphics[scale=0.1]{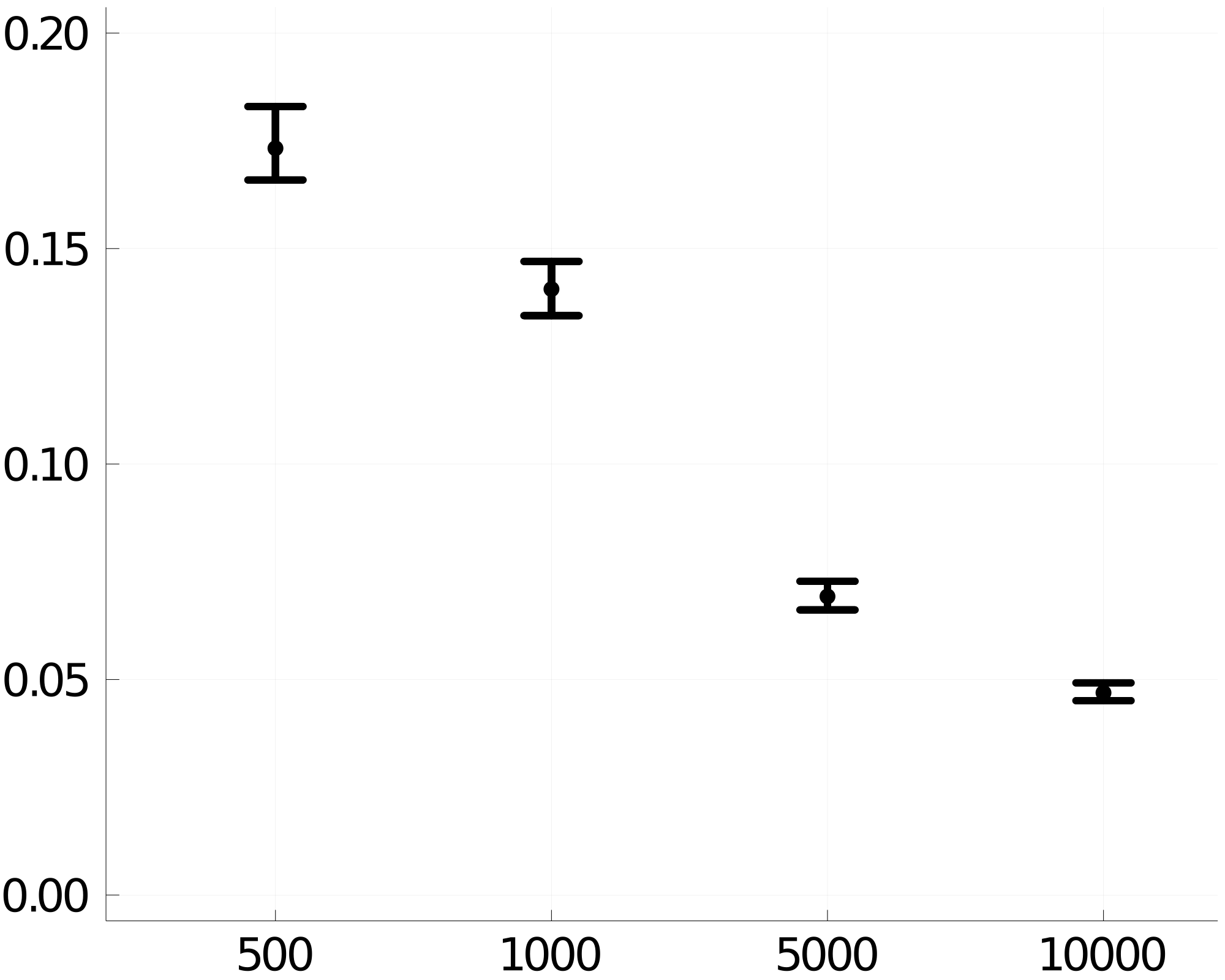} 
     }\\       
     \subfigure[]{%
            \includegraphics[scale=0.1]{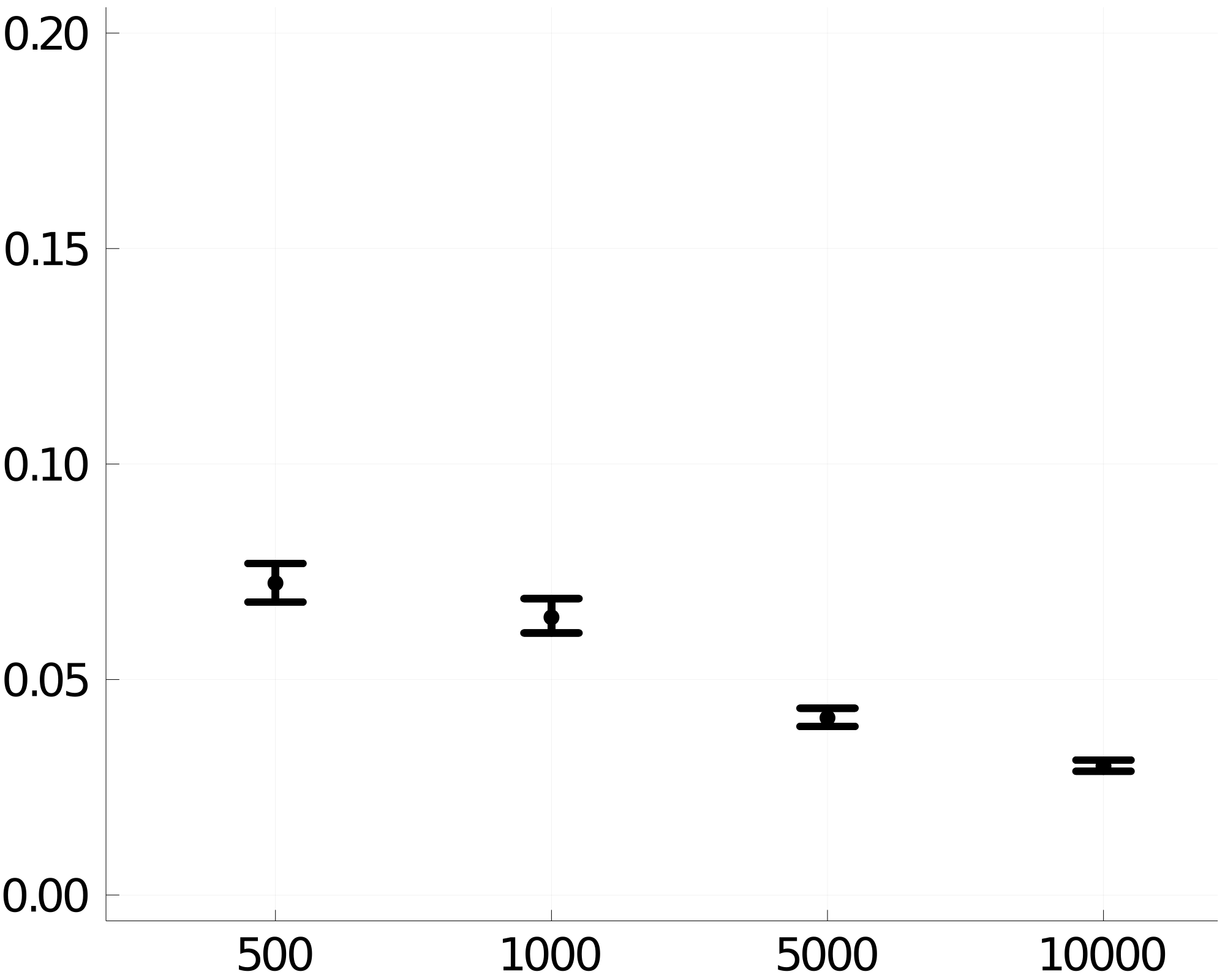}
     }\\
     \subfigure[]{%
            \includegraphics[scale=0.1]{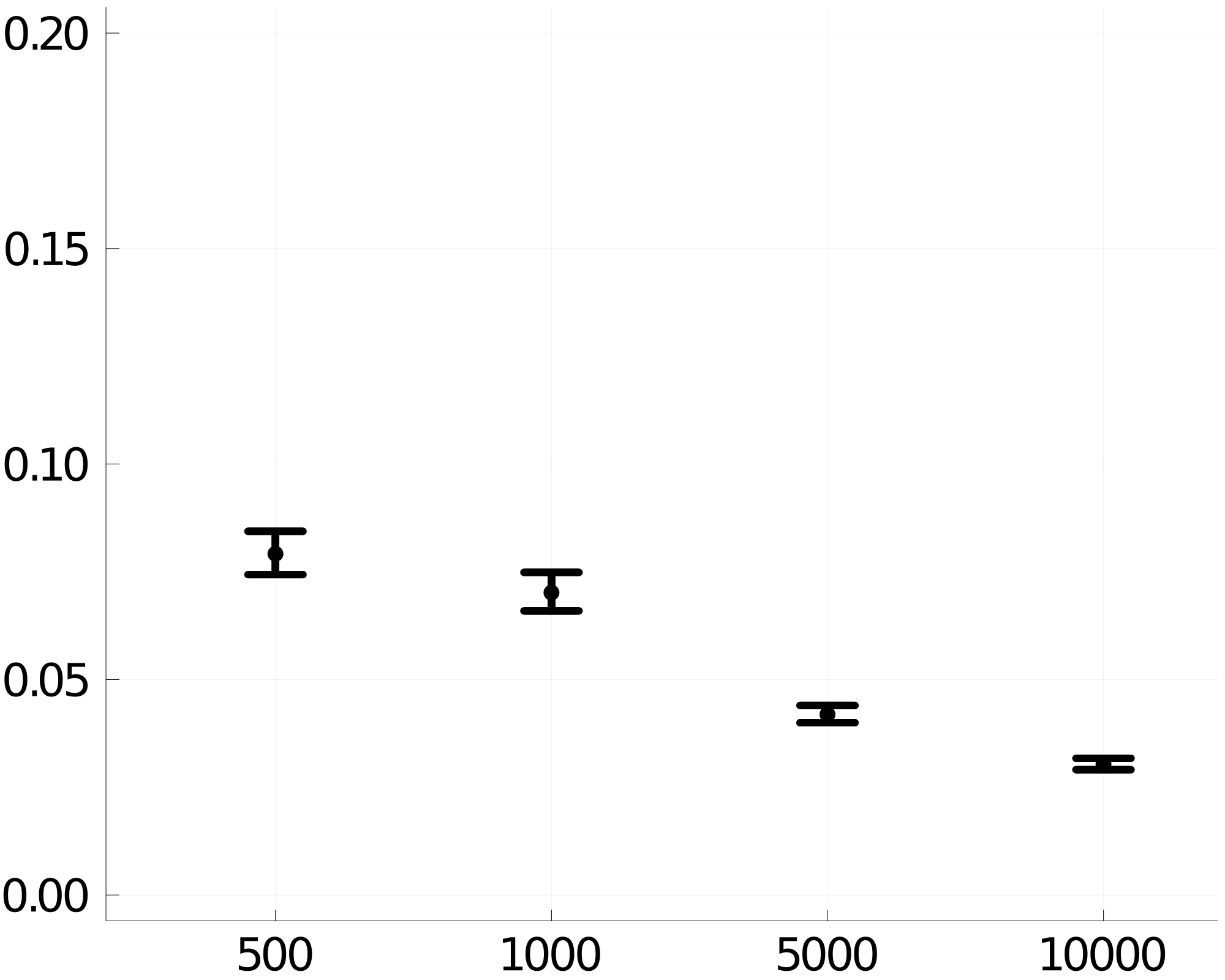}
     }       
     \caption{Posterior mean (point) and $95\%$ credibility interval (vertical bar) for the Hellinger distance to the  true joint distribution, for different sample sizes. Panel (a) displays the results for Model 1 (Clayton copula). Panel (b)  displays the results for Model 2 (Gaussian copula). Finally, panel (c) displays the results for Model 3 (a copula of a mixture of Gaussian distributions).}
        \label{resultadoshel}
    \end{centering}
\end{figure}

Figure \ref{resultadoshel} displays the posterior mean and $95\%$ credibility intervals for the Hellinger distance to the true model under the different models and sample sizes. The results show that adequate estimates for complex true models can be obtained, even for reduced sample sizes, and that when the copula model is simple, the proposed model does not overfit the data. The results also show that the posterior mean gets closer to the true model when the sample size increases and that the posterior distribution concentrates around the true model as the sample size increases.  

\subsection{Comparison with existing approaches}

We compare our proposal with the flat prior, which is one of those proposed by \cite{bnpjeffreys}. There are similarities between our model and their flat proposal. The model proposed by \cite{bnpjeffreys} is a specific case of a grid-uniform copula, restricted to two dimensions and with grids that are necessarily evenly spaced. When these conditions hold, the models differs in way the prior probability mass is assigned. The work of \cite{bnpjeffreys} focused mainly on reference priors, whereas our prior is designed to share information on neighboring sets. We compare our proposal with the flat prior proposed by \cite{bnpjeffreys} here. The flat prior of \cite{bnpjeffreys} can be thought as a limiting case of our model when $\alpha \longrightarrow 0$.

We compare the models under settings considered by  \cite{bnpjeffreys} in the evaluation of their proposal. We consider a two dimensional problem, a $6 \times 6$ grid, and a Gaussian, Gumbel and Clayton copula. We set the parameters of the different copula models such that they imply a similar association structure for the two variables. In particular, we set them such that they have the same Kendall's $\tau$, and consider $\tau = 0.05$, $0.35$, $0.50$, and $0.64$. We consider three sample sizes in the comparison, $N=30$, $N=100$, $N=400$, and $N=800$.

We performed a Monte Carlo study, considering 100 replicates for each model and sample size. For each data set we fit our proposal with the hierarchically centered prior, the ICAR correlation structure described in Section \ref{priors}, and $\alpha^\star=40$. The performance of the models was evaluated by computing the mean integrated squared error between the posterior mean and the true data generating copula model.  The results are presented in Table  \ref{table: coarsegrid}.

\begin{table}
\caption{Mean integrated squared error ($\times 10^3$) for the posterior mean of the copula function under our default default hierarchical prior (proposal) and  under the flat prior proposed by  \cite{bnpjeffreys}. The results are presented for for different true copula model, sample size ($N$), and value of Kendall's $\tau$.}
    \begin{center}
    \begin{tabular}{cc cc c cc c cc}
        \hline \hline
                     &        &  \multicolumn{2}{c}{Gaussian} &  &  \multicolumn{2}{c}{Gumbel} &  & \multicolumn{2}{c}{Clayton} \\ \cline{3-4} \cline{6-7} \cline{9-10}
        $\tau$   & $N$ &  Proposal & Flat  & &  Proposal & Flat  & & Proposal & Flat  \\\hline
        0.05 & 30  & 0.49934  &  0.86524 & & 0.42447  &  0.89990 & & 0.45682  &  0.86651  \\
        0.05 & 100 & 0.26849  &  0.35385 & & 0.26786  &  0.54728 & & 0.26364  &  0.52890  \\
        0.05 & 400 & 0.11716  &  0.18555 & & 0.11664  &  0.19051 & & 0.11903  &  0.19233  \\
        0.05 & 800 & 0.07587  &  0.10114 & & 0.07394  &  0.10226 & & 0.07109  &  0.10378  \\\hdashline
        0.35 & 30  & 0.71649  &  1.55763 & & 0.74654  &  1.60425 & & 0.70620  &  1.75942  \\
        0.35 & 100 & 0.31038  &  0.68425 & & 0.30805  &  0.65882 & & 0.31625  &  0.69027  \\
        0.35 & 400 & 0.12172  &  0.18451 & & 0.12707  &  0.19065 & & 0.12972  &  0.19401  \\
        0.35 & 800 & 0.08084  &  0.09787 & & 0.08706  &  0.10498 & & 0.08334  &  0.09943  \\\hdashline
        0.50 & 30  & 0.96715  &  2.44198 & & 1.23920  &  2.32074 & & 1.04132  &  2.43654  \\
        0.50 & 100 & 0.38696  &  0.81238 & & 0.45984  &  0.79660 & & 0.39790  &  0.79267  \\
        0.50 & 400 & 0.13393  &  0.19244 & & 0.15393  &  0.19459 & & 0.14658  &  0.18812  \\
        0.50 & 800 & 0.08496  &  0.10216 & & 0.08648  &  0.10053 & & 0.08823  &  0.10280  \\\hdashline
        0.64 & 30  & 1.26014  &  3.02553 & & 1.32796  &  2.97399 & & 1.20894  &  3.10015  \\
        0.64 & 100 & 0.47175  &  0.91925 & & 0.47628  &  0.90992 & & 0.49692  &  0.93266  \\
        0.64 & 400 & 0.14998  &  0.20586 & & 0.15408  &  0.19657 & & 0.15943  &  0.19182  \\
        0.64 & 800 & 0.09450  &  0.10742 & & 0.09590  &  0.11065 & & 0.09769  &  0.10602  \\
        \hline \hline
    \end{tabular}
    \end{center}
    \label{table: coarsegrid}
\end{table}

The results illustrate that the proposed model outperform the flat prior across the board. As expected, the biggest differences between models are observed at small sample sizes; the larger the sample size, the smaller the difference between models regardless of the association structure. Furthermore, for a given sample size, our model tends to produce better results than the flat prior as the level of association increases.

\section{Concluding remarks}

Flexible inference of copula functions  had mainly relied on partial likelihood or pseudo-likelihood methods. This approach is useful in some cases. However, they do not allow for a proper quantification of the uncertainties associated to the lack of knowledge of the marginal distributions and cannot be employed for modelling the association structure of latent variables in the context of hierarchical models. We have proposed a novel and rich family of copula functions that can overcome these problems, the class of grid-uniform copula functions. We prove that this class is dense in the space of all continuous copula functions in a Hellinger sense.  

We proposed a hierarchically centered prior distribution based on the proposed family, borrowing ideas from spatial statistics. We have described a class of transformations on grid-uniform copulas which is closed in the space of grid-uniform copula functions and that is able to span the complete space  of grid-uniform copula functions in finite number of steps, starting from any point in the space. This family of transformations, referred to as rectangle exchanges, is employed to develop an automatic MCMC algorithm for exploring the corresponding posterior distribution. We have illustrated the behavior of the proposal and compared it with the approach  proposed by \cite{bnpjeffreys}. By considering similar simulation settings to the ones considered by \cite{bnpjeffreys}, we show that our proposal outperforms their flat model 
when the posterior mean is the point estimator and mean integrated squared error is considered as a model comparison criteria.

The proposed prior model can be extended in different ways. The current proposal depends on a user-specified grid $\rho$.  The size of the grid and the location of the points may have an important influence in the resulting model. For instance, equally spaced grids can lead to over fitting of the copula function in areas where few data points are "observed". On the other hand, they can lead to under fitting in areas where more data points are "observed". The study of strategies for the estimation of the optimal size and location of the grid is the subject of ongoing research.

The proposed model suffers from the curse of dimensionality. For a sample of size $1000$, for a grip-uniform prior with a $10\times10$ grid it takes only a few seconds to generate a Markov chain of length 20,000 using the automatic MCMC algorithm and an i5 processor. We have also been able to use the proposed model in dimensions up to six. However, the implementation of the models in high dimensions and with fine grids would result in an explosion of parameters that need to be updated, which makes the implementation of this approach practically impossible. The study of marginal versions of the model, where the copula probabilities are integrated out of the model is also subject of ongoing research.  

Finally, the extension of the model to handle mixed, discrete and continuous, variables and to copula regression problems is also subject of ongoing research. 

\appendix
\section*{Acknowledgements}
N. Kuschinski's  research is supported by supported by ANID – Millennium Science Initiative Program – NCN17\_059. A. Jara's research is supported by supported by ANID – Millennium Science Initiative Program – NCN17\_059 and Fondecyt 1180640 grant.

\bibliographystyle{apalike}
\bibliography{copula}

\begin{thebibliography}{}

\bibitem[Banerjee et~al., 2014]{carbook}
Banerjee, S., Carlin, B.~P., and Gelfand, A.~E. (2014).
\newblock {\em Hierarchical modeling and analysis for spatial data}.
\newblock CRC press.

\bibitem[Chen and Dey, 1998]{hitandrun}
Chen, M.-H. and Dey, D.~K. (1998).
\newblock Bayesian modeling of correlated binary responses via scale mixture of
  multivariate normal link functions.
\newblock {\em Sankhy{\=a}: The Indian Journal of Statistics, Series A}, pages
  322--343.

\bibitem[Choro{\'s} et~al., 2010]{summaryfreqcop}
Choro{\'s}, B., Ibragimov, R., and Permiakova, E. (2010).
\newblock Copula estimation.
\newblock In {\em Copula theory and its applications}, pages 77--91. Springer.

\bibitem[Christen et~al., 2010]{twalk}
Christen, J.~A., Fox, C., et~al. (2010).
\newblock A general purpose sampling algorithm for continuous distributions
  (the t-walk).
\newblock {\em Bayesian Analysis}, 5(2):263--281.

\bibitem[Church, 2012]{asymtcop}
Church, C. (2012).
\newblock {\em The Asymmetric t-Copula with Individual Degrees of Freedom}.
\newblock PhD thesis, University of Oxford,.

\bibitem[Deheuvels, 1979]{empcop}
Deheuvels, P. (1979).
\newblock La fonction de d{\'e}pendance empirique et ses propri{\'e}t{\'e}s. un
  test non param{\'e}trique d'ind{\'e}pendance.
\newblock {\em Bulletins de l'Acad{\'e}mie Royale de Belgique}, 65(1):274--292.

\bibitem[Eilers and Marx, 1996]{eilers1996}
Eilers, P. H.~C. and Marx, B.~D. (1996).
\newblock Flexible smoothing with {B}-splines and penalties.
\newblock {\em Statist. Sci.}, 11(2):89--121.

\bibitem[Faugeras, 2013]{sklarthm}
Faugeras, O.~P. (2013).
\newblock Sklar’s theorem derived using probabilistic continuation and two
  consistency results.
\newblock {\em Journal of Multivariate Analysis}, 122.

\bibitem[Genest et~al., 1995]{semiparamfreq}
Genest, C., Ghoudi, K., and Rivest, L.-P. (1995).
\newblock A semiparametric estimation procedure of dependence parameters in
  multivariate families of distributions.
\newblock {\em Biometrika}, 82(3):543--552.

\bibitem[Genest and Ne{\v{s}}lehov{\'a}, 2007]{countcops}
Genest, C. and Ne{\v{s}}lehov{\'a}, J. (2007).
\newblock A primer on copulas for count data.
\newblock {\em ASTIN Bulletin: The Journal of the IAA}, 37(2):475--515.

\bibitem[Genest et~al., 2011]{archimcop}
Genest, C., Ne{\v{s}}lehov{\'a}, J., and Ziegel, J. (2011).
\newblock Inference in multivariate archimedean copula models.
\newblock {\em Test}, 20(2):223.

\bibitem[Guillotte and Perron, 2012]{bnpjeffreys}
Guillotte, S. and Perron, F. (2012).
\newblock Bayesian estimation of a bivariate copula using the jeffreys prior.
\newblock {\em Bernoulli}, 18(2):496--519.

\bibitem[Joe, 2014]{librocop1}
Joe, H. (2014).
\newblock {\em Dependence modeling with copulas}.
\newblock CRC press, Boca Raton.

\bibitem[Kaewsompong et~al., 2020]{bayesarchim}
Kaewsompong, N., Maneejuk, P., and Yamaka, W. (2020).
\newblock Bayesian estimation of archimedean copula-based {SUR} quantile
  models.
\newblock {\em Complexity}, 2020:1--15.

\bibitem[McNeil and Ne{\v{s}}lehov{\'{a}}, 2009]{archimcop1}
McNeil, A.~J. and Ne{\v{s}}lehov{\'{a}}, J. (2009).
\newblock Multivariate archimedean copulas, d-monotone functions and
  $\ell$1-norm symmetric distributions.
\newblock {\em The Annals of Statistics}, 37(5B):3059--3097.

\bibitem[Mukhopadhyay and Parzen, 2020]{freqnpcop}
Mukhopadhyay, S. and Parzen, E. (2020).
\newblock Nonparametric universal copula modeling.
\newblock {\em Applied Stochastic Models in Business and Industry},
  36(1):77--94.

\bibitem[Müller et~al., 2015]{bnpbook}
Müller, P., Quintana, F., Jara, A., and Hanson, T. (2015).
\newblock {\em {Bayesian Nonparametric Data Analysis}}.
\newblock Springer, New York, USA.

\bibitem[Nelsen, 2007]{librocop}
Nelsen, R.~B. (2007).
\newblock {\em An introduction to copulas}.
\newblock Springer Science \& Business Media, New York.

\bibitem[Ning and Shephard, 2018]{polyacop}
Ning, S. and Shephard, N. (2018).
\newblock A nonparametric bayesian approach to copula estimation.
\newblock {\em Journal of Statistical Computation and Simulation},
  88(6):1081--1105.

\bibitem[Pitt et~al., 2006]{bayesgausscop}
Pitt, M., Chan, D., and Kohn, R. (2006).
\newblock Efficient bayesian inference for gaussian copula regression models.
\newblock {\em Biometrika}, 93(3):537--554.

\bibitem[Robert and Casella, 2013]{mcmc}
Robert, C. and Casella, G. (2013).
\newblock {\em Monte Carlo statistical methods}.
\newblock Springer Science \& Business Media, New York.

\bibitem[Schmidt and Nobre, 2014]{carprocs}
Schmidt, A.~M. and Nobre, W.~S. (2014).
\newblock Conditional autoregressive (car) model.
\newblock {\em Wiley StatsRef: Statistics Reference Online}, pages 1--11.

\bibitem[Sklar, 1959]{sklarthm1}
Sklar, M. (1959).
\newblock Fonctions de repartition an dimensions et leurs marges.
\newblock {\em Publ. inst. statist. univ. Paris}, 8:229--231.

\bibitem[Smith, 2011]{summarycop}
Smith, M.~S. (2011).
\newblock Bayesian approaches to copula modelling.
\newblock {\em arXiv preprint arXiv:1112.4204}.

\bibitem[Wu et~al., 2013a]{gausmixcop}
Wu, J., Wang, X., and Walker, S.~G. (2013a).
\newblock Bayesian nonparametric estimation of a copula.
\newblock {\em Journal of Statistical Computation and Simulation},
  85(1):103--116.

\bibitem[Wu et~al., 2013b]{skewmixcop}
Wu, J., Wang, X., and Walker, S.~G. (2013b).
\newblock Bayesian nonparametric inference for a multivariate copula function.
\newblock {\em Methodology and Computing in Applied Probability},
  16(3):747--763.

\bibitem[Yang et~al., 2019]{discretecops}
Yang, L., Frees, E.~W., and Zhang, Z. (2019).
\newblock Nonparametric estimation of copula regression models with discrete
  outcomes.
\newblock {\em Journal of the American Statistical Association},
  115(530):707--720.

\end{thebibliography}

\end{document}